%% file: main.tex
\documentclass[11pt]{article}

\usepackage{amsmath}
\usepackage{amssymb}
\usepackage{amsthm} \usepackage[linesnumbered,ruled,vlined]{algorithm2e} 
\usepackage{mathtools}
\usepackage{graphicx} % support the \includegraphics command and options
\usepackage{array} % for better arrays (eg matrices) in maths

\usepackage[usenames,dvipsnames]{xcolor}
\usepackage[ruled]{algorithm2e}

\usepackage{tcolorbox}
\tcbuselibrary{skins,breakable}
\tcbset{enhanced jigsaw}

\usepackage[compact]{titlesec}

\definecolor{DarkRed}{rgb}{0.5,0.1,0.1}
\definecolor{DarkBlue}{rgb}{0.1,0.1,0.5}

\usepackage{nameref}
\definecolor{ForestGreen}{rgb}{0.1333,0.5451,0.1333}
\definecolor{Red}{rgb}{0.9,0,0}
\usepackage[linktocpage=true,
	pagebackref=true,colorlinks,
	linkcolor=DarkRed,citecolor=ForestGreen,
	bookmarks,bookmarksopen,bookmarksnumbered]
	{hyperref}

\usepackage{bm}
\usepackage{url}
\usepackage{xspace}
\usepackage[mathscr]{euscript}

\usepackage{mdframed}

\usepackage[noend]{algpseudocode}
\makeatletter
\def\BState{\State\hskip-\ALG@thistlm}
\makeatother

\usepackage{cite}
\usepackage{enumitem}

\usepackage[margin=1in]{geometry}

\usepackage{tikz}
\usetikzlibrary{arrows}
\usetikzlibrary{arrows.meta}
\usetikzlibrary{shapes}
\usetikzlibrary{backgrounds}
\usetikzlibrary{positioning}
\usetikzlibrary{decorations.markings}
\usetikzlibrary{patterns}
\usetikzlibrary{calc}
\usetikzlibrary{fit}
\usetikzlibrary{snakes}
\usetikzlibrary{shadows.blur}
\usetikzlibrary{petri,decorations.markings}

\usepackage{caption}
\usepackage{subcaption}

\newcommand{\expp}[2]{\mathbb{E}_{#1}\bracket{#2}}

\newtheorem{theorem}{Theorem}
\newtheorem{lemma}{Lemma}[section]
\newtheorem{proposition}[lemma]{Proposition}

\newtheorem{claim}[lemma]{Claim}

\newtheorem{problem}{Problem}
\newtheorem*{lemma*}{Lemma}
\newtheorem{definition}{Definition}
\newcommand*\samethanks[1][\value{footnote}]{\footnotemark[#1]}

\input{macros}

\title{Near-Perfect Recovery in the One-Dimensional Latent Space Model} 

\author{Yu Chen\thanks{Department of Computer and Information Science, University of Pennsylvania. Email: {{\small {\tt \{chenyu2,kannan,sanjeev\}@cis.upenn.edu.}}}}
\and Sampath Kannan\samethanks
\and Sanjeev Khanna\samethanks
}

\begin{document}

\maketitle

\input{abstract}

\input{intro}

\input{math}

\input{order}

\input{order_alg}

\input{order-exp}

\input{linear}

\input{position}

\input{lb}

\input{ub}

\input{exper}

\input{conclusions}

\bibliographystyle{abbrv}
\bibliography{general}

\end{document}

%% file: macros.tex
\newcommand{\toShrink}{-.20cm}
\newcommand{\toShrinkEnu}{-.2cm}

\newcommand{\tvd}[2]{\ensuremath{\norm{#1 - #2}_{tvd}}}
\renewcommand{\tvd}[2]{\ensuremath{\Delta_{\textnormal{\texttt{TV}}}(#1,#2)}}

\newcommand{\eps}{\ensuremath{\varepsilon}}

\newcommand{\bracket}[1]{\left[#1\right]}
\newcommand{\paren}[1]{\ensuremath{\left(#1\right)}\xspace}
\newcommand{\card}[1]{\left\vert{#1}\right\vert}

\newcommand{\kl}[2]{\ensuremath{\ID_{KL}(#1 \parallel #2)}}

\newcommand{\ID}{\ensuremath{\mathbb{D}}}

\newcommand{\norm}[1]{\ensuremath{\|#1\|}}

\newcommand{\prob}[1]{\Pr\paren{#1}}
\newcommand{\expect}[1]{\Exp\bracket{#1}}

\DeclareMathOperator*{\Exp}{\ensuremath{{\mathbb{E}}}}
\DeclareMathOperator*{\Prob}{\ensuremath{\textnormal{Pr}}}
\renewcommand{\Pr}{\Prob}

\newenvironment{tbox}{\begin{tcolorbox}[
		enlarge top by=5pt,
		enlarge bottom by=5pt,
		 breakable,
		 boxsep=0pt,
                  left=4pt,
                  right=4pt,
                  top=10pt,
                  boxrule=1pt,toprule=1pt,
                  colback=white,
                  arc=-1pt,
                  ]%%
	}
{\end{tcolorbox}}

%% file: abstract.tex
\begin{abstract}

Suppose a graph $G$ is stochastically created by uniformly sampling vertices along a line segment and connecting each pair of vertices with a probability that is a known decreasing function of their distance. We ask if it is possible to reconstruct the actual positions of the vertices in $G$ by only observing the generated unlabeled graph. We study this question for two natural edge probability functions --- one where the probability of an edge decays exponentially with the distance and another where this probability decays only linearly. We initiate our study with the weaker goal of recovering only the order in which vertices appear on the line segment.
For a segment of length $n$ and a precision parameter $\delta$, we show that for both exponential and linear decay edge probability functions, there is an efficient algorithm that correctly recovers (up to reflection symmetry) the order of all vertices that are at least $\delta$ apart, using only $\tilde{O}(\frac{n}{\delta ^ 2})$ samples (vertices). Building on this result, we then show that $O(\frac{n^2 \log n}{\delta ^2})$ vertices (samples) are sufficient to additionally recover the location of each vertex on the line to within a precision of $\delta$. 
We complement this result with an $\Omega (\frac{n^{1.5}}{\delta})$ lower bound on samples needed for reconstructing positions (even by a computationally unbounded algorithm), showing that the task of recovering positions is information-theoretically harder than recovering the order. We give experimental results showing that our algorithm recovers the positions of almost all points with high accuracy.
\end{abstract}

%% file: intro.tex
\section{Introduction}

Large graphs arise naturally in modeling many scenarios in social interaction, natural language processing, image processing, and recommendation systems. Nodes in these graphs
represent individual entities such as people, genes, or pixels and edges represent relationships between them.  A natural goal in analyzing such graphs is to partition the nodes into a small number of sets in such a way that two nodes in the same set `behave similarly' in terms of their interaction. Algorithms for finding such \textit{communities} are analyzed on synthetic data generated by a stochastic model. The \textit{stochastic block model} or \textit{planted cluster} model is a commonly used generative model. This model is parametrized by $(n, k ,\pi, P)$ where $n$ is the number of vertices, $k$ is the number of clusters, $\pi$ is a $k$-vector of probabilities summing to 1, and $P$ is a $k \times k$ matrix. 
The cluster that a vertex belongs to is chosen independently of other vertices according to $\pi$. For any two vertices $u$ and $v$ in clusters $i$ and $j $ respectively, the probability of an edge between $u$ and $v$ is $P[i,j]$. Much work has been done in this model to understand the information-theoretic and computational limits for achieving \textit{exact, partial} and \textit{weak} recovery. For a detailed discussion of the model, its motivation, different notions of recovery, and positive and negative results, see the excellent survey by Abbe~\cite{abbe2018community}.

The stochastic block model is based on the assumption that the entities involved can be neatly categorized into a small number of classes, and  membership in a class is the sole determinant of how an entity interacts with others.  For example, in this model, we could regard people's political persuasion as being binary -- say, liberal or conservative in the United States -- and posit that there is a  certain probability for edges connecting two  conservatives or two liberals, and a different probability for an edge connecting a liberal to a conservative. Many real situations are more complex. For example, the probability of an edge between two nodes in a social network might be a function of many different \textit{attributes} of these nodes, each of which can be discrete or continuous-valued. 

Other variants of the stochastic block model have been proposed recently.~\cite{deshpande2018contextual,ke2018information}. However, these models all share many features with the stochastic block model, and in particular, assume that objects only belong to one of a small number of clusters, with a clear difference in probabilities between intra-cluster and inter-cluster edges. Just like the stochastic block model, these models do not capture some aspects of the real situation in which relationship graphs arise.

In this paper, we study similar recovery problems in a different model called the {\em latent space model}. In this model, we think of nodes as points in a metric space, and let edges be independently sampled with probabilities that are a decreasing function of the distance between the endpoints. Given a large graph generated according to this model, we seek to find (approximate) locations of each node or entity in the metric space. The latent space model can be seen as a generalization of the stochastic block model, by letting the  points in the same cluster be at distance 0 from each other, and points in different clusters be at distance 1. In fact, an intermediate model between the stochastic block model and our model consists of a metric space with a finite number of points (or clusters), where each entity is located at one of these points. If we can find good enough approximations for the location of each node in the metric space, we will exactly identify cluster membership in these finite and discrete metric spaces.

The latent space model was first introduced by Hoff et al.~\cite{hoff2002latent} and extended by Handcock et al. ~\cite{handcock2007model}. This model has been applied to political relationships~\cite{hoff2004modeling,ng2018modeling} and social networks~\cite{fisher2019social}. Previous work on this model has been focused on heuristic approaches to finding the maximum likelihood latent positions and empirical evaluations of these approaches~\cite{handcock2007model,hoff2002latent,raftery2012fast}. Recently, Ke and Honorio~\cite{ke2018information,ke2019exact} studied a particular case of the latent space model where each point belongs to one of two communities, and the points in the same community are close to each other.

We study the basic version of the latent space model, where the nodes are uniformly sampled on a segment. We consider both the problem of recovering the order of the nodes and the problem of recovering the positions of the nodes. For this simple setting our focus is on designing algorithms with provable guarantees on number of samples needed, running time, and quality of approximation. 
 While maximum likelihood methods are statistically consistent and converge to the right model in the limit, there are no proofs in the literature about the convergence rate of maximum likelihood heuristics that have been proposed for our model.
%Our goal of finding approximate positions for the vertices is also different from the goal of finding the most likely positions. 

The work of Sarkar {\em et. al}~\cite{DBLP:conf/ijcai/SarkarCM11} considers a setting that is somewhat similar to ours. They focus on the problem of estimating the distance between a given pair of nodes in a $d$-dimensional latent space, based on the observed graph. However, in their setting, the edges obey a threshold behavior where any pair of nodes has an edge iff they are within a specified threshold distance. Thus once the node positions are fixed, the resulting graph is deterministic. In contrast, in our setting, even when the node positions are fixed, the resulting graph has high entropy as well contains a mixture of short-range and long-range edges, making the reconstruction problem distinctly more challenging even in the one-dimensional case that we consider here.

In statistical mechanics and probability theory, models such as the latent space model have been studied under the name \textit{long-range percolation models}~\cite{aizenman1986discontinuity,coppersmith2002diameter,newman1986one,schulman1983long}. Most of the work in these disciplines is focused on the problem of understanding structural properties of the graphs that arise, rather than algorithmic reconstruction of the locations of entities. 
Our paper takes a first step in designing and analyzing efficient algorithms for this reconstruction. We focus here on reconstruction in a one-dimensional metric space, namely, the real interval $[0,n]$. We assume that entities are uniformly sampled (with sufficient density) from this metric space. We also restrict attention to specific types of edge probability functions - exponentially decaying functions and linearly decaying functions. In other words, if $d$ is the distance between points $u$ and $v$, we consider a model where the probability of an edge is $c e^{-d}$ and another model where the probability of an edge is
$\frac{c}{d + 1}$, in both cases for a constant $0<c\le 1$. 

\iffalse
In the stochastic block model, where the problem is to identify the cluster to which each entity belongs, 3 types of recovery are considered: \textbf{Exact recovery} where the goal is to identify the cluster membership of every entity with probability close to 1, \textbf{Almost exact recovery}, where the goal is to identify the cluster memberships of all but a vanishingly small set of entities with probability close to 1, and \textbf{Partial recovery} where the cluster memberships of a constant fraction of the points is determined with probability close to 1. \textbf{Weak recovery} is the weakest possible kind of partial recovery, where the fraction of points correctly identified is bounded away from the trivial threshold, which is achieved by an algorithm that ignores the input and randomly guesses the cluster to which each point belongs. In our model, we cannot hope to find the exact location of any point given the finite number of nodes and the fact that locations are only random variables estimated from a stochastic process. Thus, at best we can only hope to locate each node within an interval of some width $\delta$, that depends on the density with which nodes are sampled. With this caveat, we can equivalently define exact, almost exact, partial, and weak recovery. Specifically, in exact (resp. almost exact, partial, weak) recovery, the goal is to approximate the order or the positions of all (resp. almost all, a constant fraction, a non-trivial fraction) of the entities within some constant error. 
\fi

In the standard stochastic model a distinction is made between fundamental (information-theoretic) limits and (efficient) computational limits for each kind of recovery and bounds for each of them are pretty tightly pinned down. Specifically, the information-theoretic bounds are based on the separation needed between intra-cluster edge probabilities and inter-cluster probabilities. Since our edge probabilities are continuous functions of distance, we cannot hope to show these kinds of bounds. Instead, we give upper and lower bounds for how densely entities must be sampled in order to efficiently recover their approximate order. 
Since these bounds are essentially tight, and the upper bound is by an efficient algorithm, they are both information-theoretic and computational.

\subsection{Problem Statement and Results}
We consider the following scenario: On the segment $[0,n]$ $m$ points, say $v_1,v_2,\dots,v_m$, are uniformly sampled. Let $x_i$ be the location of $v_i$, and let $X=(x_1,x_2,\dots,x_m)$ be the location vector. A random graph $G$ is constructed with this vertex set;  edges are sampled independently as follows: for any pair of vertices $v_i$ and $v_j$, an edge exists between them with probability $c \cdot f(\card{x_i-x_j})$, where $0<c \le 1$ and $f$ is some monotone decreasing function such that $f(0)=1$ and $\lim_{x \rightarrow \infty} f(x)=0$. For such a graph $G$ and a position vector $X$, denote by $P_X(G)$ the likelihood of $G$ given $X$, i.e. $P_X(G) = \prod_{(i,j)\in G} c \cdot f(\card{x_i-x_j}) \cdot \prod_{(i,j)\notin G} (1-c \cdot f(\card{x_i-x_j}))$.

Our goal is to design an algorithm that takes as input the (unlabeled) graph $G$, and a constant $\delta$, and outputs  a vector $(\hat{x}_1,\hat{x}_2,\dots,\hat{x}_m)$ which is a ``recovery'' of the location of each point. We consider two distinct notions of recovery: (1) recovering the order, by which we mean that for any pair of $i$ and $j$ such that $x_i-x_j > \delta$, $\hat{x}_i > \hat{x}_j$ with high probability; (2) recovering the location, by which we mean that  for any $i$, $\card{x_i-\hat{x}_i}<\delta$ with high probability. We study both these problems for two natural choices of $f$, namely, the exponential decay function $f(x)=e^{-x}$, and the linear decay function $f(x) = \frac{1}{x+1}$. 

For the problem of recovering the order to within any specified precision $\delta$, we show that it suffices to sample $m = \tilde{O}(\frac{n}{\delta^2})$ points. Notice that $\Omega(n \log n)$ points are necessary, since otherwise $G$ will have isolated vertices with high probability, and it is information-theoretically infeasible to determine the relative order of two isolated vertices no matter how far apart. At a high-level, our algorithms employ the following general approach. First, for each pair of vertices, we use the number of common neighbors to approximate the distance between them. Of course the greater the number of common neighbors between two nodes, the smaller we expect their distance to be. But we need to precisely quantify the range of distances that can be sufficiently accurately reconstructed using a coarse measure such as the number of common neighbors. We prove bounds for this range under both exponential decay and linear decay models. We then use this information to determine spatial relationships between vertices, and recover a global order. 

For the problem of recovering the location, we focus on the case $c=1$. Building on our algorithm for recovering the order, we can show that with  $m=O(n^2\log n/\delta^2)$ samples, it is possible to recover locations of the points to within precision $\delta$. We also show that the sample complexity of recovering positions is inherently much more than the sample complexity for recovering the order. Specifically,
 for any $m = o(n^{1.5}/\delta)$, we give two location vectors $X^1$ and $X^2$ such that $\norm{X^1-X^2}_{\infty} > \delta$ and prove that it is impossible to distinguish these two vectors with large constant probability given a random graph $G$ generated in accordance with one of these two vectors. This shows that  $\Omega(n^{1.5}/\delta)$ points are necessary to recover locations. Matching this, given $m=\Omega(n^{1.5}\log n/\delta)$ samples, we prove that we can distinguish between any two location vector $X^1$ and $X^2$ such that $\norm{X^1-X^2}_\infty > \delta$.  Note that the $\tilde{O}(n^{1.5})$ upper bound refers to the problem of distinguishing two position vectors. The best upper bound we can prove for recovering position is still $\tilde{O}(n^2)$.
 
Finally, we analyze the accuracy of our recovery algorithms on synthetically generated datasets, and show that consistent with our theoretical results, we are able to reconstruct the order and positions of the underlying point set to an increasingly high precision as the sample size increases.

\smallskip
\noindent
\subsection*{Organization} The remainder of the paper is organized as follows. 
In Section~\ref{sec:math}, we give some math results which we will use in our paper. In Section~\ref{sec:order}, we present and analyze our algorithm for recovering the order of vertices for both the exponential decay function and the linear decay function. In Section~\ref{sec:position}, we show that we can recover approximate positions of each vertex in both models. We also establish our lower bound on the number of samples needed for this task. We present our empirical results in Section~\ref{empirical}. Finally, in Section~\ref{sec:conclusions} we briefly discuss  the larger context for our problem and open problems.

%% file: math.tex
\section{Math Tools} \label{sec:math}

\subsection{Basic Math Inequalities}

In this section, we prove some math results we used.

\begin{proposition} \label{prop:linear}
    Suppose four different numbers $a$, $a'$, $b$, $b'$, $\eps$ satisfy that $0\le \eps < 1/2$, $\card{a-a'}<\eps a$, $\card{b-b'}<\eps b$, and $4<a<b$, then $\card{\frac{\log b - \log a}{b-a} - \frac{\log b'- \log a'}{b'-a'}} < \eps$ 
\end{proposition}

\begin{proof}
    For any positive numbers $i,j$, let $g(i,j) = \frac{\log i -\log j}{j - i}$. Then $g(i,j) = \int_i^j \frac{1}{x} dx$, which means $g(i,j)$ is between $\frac{1}{i}$ and $\frac{1}{j}$.

    We first prove $\card{g(a,b)-g(a',b)} < \frac{\eps}{2}$, and with the same argument, $\card{g(a',b)-g(a',b')} < \frac{\eps}{2}$, which together imply the proposition.

    \textbf{Case 1:} $a' < a < b$. $g(a',b) = \frac{b-a}{b-a'}g(a,b) + \frac{a-a'}{b-a'}g(a,a')$, which means $\card{g(a',b)-g(a,b)} = \frac{a-a'}{b-a'} \card{g(a,a')-g(a,b)} < \frac{a-a'}{b-a'} (\frac{1}{a'}-\frac{1}{b}) = \frac{a-a'}{a'b} < \frac{2\eps}{b} < \frac{\eps}{2}$. 

    \textbf{Case 2:} $a < a' < b$. $g(a,b) = \frac{b-a'}{b-a}g(a',b) + \frac{a'-a}{b-a}g(a',a)$, which means $\card{g(a',b)-g(a,b)} = \frac{a-a'}{b-a} \card{g(a,a')-g(a',b)} < \frac{a'-a}{b-a} (\frac{1}{a} - \frac{1}{b} = \frac{a'-a}{ab} < \frac{\eps}{b} < \frac{\eps}{4}$. 

    \textbf{Case 3:} $a < b < a'$, $\card{g(a,a')-g(b,a')} < \frac{1}{a}-\frac{1}{a'} < \frac{\eps}{a'} < \frac{\eps}{4}$.
\end{proof}

\begin{proposition} \label{cla:1-y}
    If $0<x$, $x+x^2/2 < \log (1-x)$; if $x<0.5$, $\log (1-x) < x+x^2$.
\end{proposition}

\begin{proof}
    The Taylor expansion of $\log (1-x)$ is 
    $$
    -\log (1-x) = \sum_{k=1}^{\infty} \frac{x^k}{k} > x+x^2/2
    $$
    The inequality holds because $x>0$.
    On the other hand,
    $$
    \sum_{k=1}^{\infty} \frac{x^k}{k} < x + \frac{1}{2}\sum_{k=2}^{\infty}x^k < x+x^2
    $$
    since $x<0.5$.
\end{proof}

\begin{proposition} \label{cla:log}
    For any $0<x'\le x$, $\frac{e^{-x}(e^{x'}-1)}{1-e^{-x}} \le \frac{x'}{x}$.
\end{proposition}
\begin{proof}
    Let $\eps=\frac{x'}{x}$, to prove the proposition, we only need to prove that for any $0<\eps \le 1$, $\frac{e^{-x}(e^{\eps x}-1)}{1-e^{-x}}<\eps$, which is equivalent to proving that 
    $e^{(\eps-1)x}-(1-\eps)e^{-x}<\eps$
    
    Let $f_{\eps}(x)$ be the LHS, $f_{\eps}(0)=\eps$. The derivative 
    $f'_{\eps}(x)=(\eps-1)e^{(\eps-1)x}-(\eps-1)e^{-x}<0$ when $x>0$, so $f_{\eps}(x)<\eps$ when $x>0$.
\end{proof}

\begin{proposition} \label{cla:log2}
    For any $0<x' \le x$, $\frac{e^{-x}(1-e^{-x'})}{1-e^{-x}} \le \frac{x'}{x}$.
\end{proposition}
\begin{proof}
    Let $\eps=\frac{x'}{x}$, to prove the proposition, we only need to prove that for any $\eps>0$, $\frac{e^{-x}(1-e^{-\eps x})}{1-e^{-x}}<\eps$, which is equivalent to prove that
    $$
    (1+\eps)e^{-x}-e^{-(\eps+1)x}<\eps
    $$
    Let $f_{\eps}(x)$ be the LHS, $f_{\eps}(0)=\eps$, and the derivative 
    $$f'_{\eps}(x)=-(\eps+1)e^{-x} + (\eps+1)e^{-(\eps+1)x}<0$$ when $x>0$, so $f_{\eps}(x)<\eps$ when $x>0$.
\end{proof}

\begin{proposition} \label{cla:log3}
    For any $x'>x$, $\frac{1-e^{-x'}}{1-e^{-x}} < \frac{x'}{x}$.
\end{proposition}
\begin{proof}
    Let $\eps=\frac{x'}{x}$, to prove the proposition, we only need to prove that for any $\eps>1$, $\frac{1-e^{-\eps x}}{1-e^{-x}}<\eps$, which is equivalent to prove that
    $$
    e^{-\eps x}-\eps e^{-x} + \eps - 1 > 0
    $$
    Let $f_{\eps}(x)$ be the LHS, $f_{\eps}(0)=0$, and the derivative 
    $$f'_{\eps}(x)=-\eps e^{-\eps x} + \eps e^{-x} > 0$$ when $x>0$ and $\eps > 1$, so $f_{\eps}(x)>0$ when $x>0$.
\end{proof}

\subsection{Sub-exponential Variables and Bernstein Bound}

In this section, we review the concept of sub-exponential variables and Bernstein bound.

\begin{definition} [Sub-exponential Variables] \label{def:se}
    A random variable $X$ with mean $\mu$ is sub-exponential with parameters $(\sigma,b)$ if for any $\lambda$ with $\card{\lambda}<1/b$,
    $$
    \expect{e^{\lambda(X-\mu)}}\le e^{\sigma^2\lambda^2/2}
    $$
\end{definition}

The following result is a common technique for proving sub-exponential.

\begin{proposition} \label{cla:subexp}
    For any random variable $X$ with mean $\mu$ and any number $\lambda$, $\expect{e^{\lambda(X-\mu)}} < \expect{e^{\frac{\lambda^2(X-X')^2}{2}}}$ where $X'$ is a random variable which is independent and identical to $X$.
\end{proposition}

\begin{proof}
    $$\expp{X}{e^{\lambda(X-\mu)}} = \expp{X}{e^{\lambda (X - \expp{X'}{X'})}} \le \expp{X,X'}{e^{\lambda(X-X')}}$$
    The second inequality is due to Jensen’s inequality. Let $\eps$ be a random variable taking value on $\pm 1$ with probability half on both values. Since $X$ and $X'$ are identical, $\eps(X-X')$ and $X-X'$ are identical. So we have
    $$
    \expp{X,X'}{e^{\lambda(X-X')}} = \expp{X,X'}{\expp{\eps}{e^{\eps\lambda(X-X')}}}
    $$
    On the other hand, for any number $Y$, 
    \begin{align*}
        \expp{\eps}{e^{\eps Y}} &= \frac{1}{2}(e^Y+e^{-Y}) 
        = \frac{1}{2}\sum_{k=1}^{\infty} (\frac{Y^k}{k!} + \frac{(-Y)^k}{k!}) \\
        &= \sum_{k=1}^{\infty} (\frac{Y^{2k}}{(2k)!}) 
        < \sum_{k=1}^{\infty} (\frac{Y^{2k}}{2^k k!}) 
        = e^{Y^2/2}
    \end{align*}
    So $\expp{X,X'}{\expp{\eps}{e^{\eps\lambda(X-X')}}} < \expp{X,X'}{e^{\frac{\lambda^2(X-X')^2}{2}}}$
\end{proof}

\begin{proposition}[Bernstein bound \cite{bernstein1924modification}] \label{prop:Hof}
    Let $X_1,X_2,\dots,X_n$ be independent random variables, where $X_i$ is sub-exponential random variable with mean $\mu_i$ and sub-exponential parameter $(\sigma_i,b_i)$.   $$
    \prob{\card{\sum_{i=1}^n (X_i-\mu_i)} \ge t} \le \begin{cases} 
                                                            2 e^{-\frac{t^2}{2\sigma_{\star}^2}} & \text{ for } 0 \le t \le \frac{\sigma_{\star}}{b} \\
                                                            2 e^{-\frac{t}{2b_{\star}}} & \text{ for } t>\frac{\sigma_{\star}}{b}
                                                      \end{cases}
    $$
    where $\sigma_{\star}^2 = \sum_{i=1}^n \sigma^2_i$ and $b_{\star} = \max_{i=1}^n b_i$
\end{proposition}

\iffalse

\begin{definition} [L{\'e}vy Concentration Function \cite{kolmogorov1958proprietes}] \label{def:levy}
    Given a random variable $X$ and a number $t$, the L{\'e}vy Concentration function $Q_X(t)$ is defined as
    $$Q_X(t)=\sup_{a\in \mathbb{R}}\prob{|X-a|<t}$$
\end{definition}

\begin{proposition}[Kolmogorov-Rogozin Inequality \cite{rogozin1961estimate}] \label{prop:KR}
    Let $X_1,X_2,\dots,X_n$ be independent random variables and let $X=X_1+X_2+\dots+X_n$. Then for any $t>0$ and any $0<t_i<t$, we have
    $$
    Q_X(t)\le 100 \frac{t}{\sqrt{\sum_{i=1}^n t_i^2(1-Q_{X_i}(t_i))}}
    $$
\end{proposition}

\fi

%% file: order.tex
\section{Recovering the Order} \label{sec:order}

We start by proving a simple statement --- that with enough samples, each segment of length $\delta$ has at least one vertex. Throughout the paper, whenever we say $1 - o(1)$, we mean $1 - 1/poly(n)$.

\begin{lemma} \label{lem:uniform}
If $m>\frac{8n\log n}{\delta^2}$ and $\delta<1$, with probability $1-o(1)$, for any non-negative integer $i$, the interval $[\frac{i\delta}{2},\frac{(i+1)\delta}{2}]$ on the segment $[0,n]$ has at least one point.
\end{lemma}

\begin{proof}
Since $\log (\frac{1}{\delta}) < \frac{1}{\delta}-1$, $m>\frac{8n \log n + 8n \log n \log (\frac{1}{\delta})}{\delta} > \frac{8n\log (\frac{n}{\delta})}{\delta}$. For any such segment, the probability that there is no point on it is $(1-\frac{\delta}{2n})^m < e^{-\frac{m \delta}{4 n}} = o(\frac{\delta}{n})$. The assertion follows by using the union bound  over all segments.
%\sampath{But there are $\frac{n}{\delta}$ such segments... so union bound doesn't work.  Do we need a factor $\log (1/\delta)$ more samples?}
\end{proof}

We will also need the following simple proposition directly implied by Chernoff bound.

\begin{proposition} \label{sam}
    Let $X=x_1+x_2+\dots+x_{m}$ be the sum of $m$ i.i.d Bernoulli samples with probability $\frac{c \cdot A}{n}$. Let $\hat{A} = \frac{Xn}{cm}$. Then the probability that $\card{\hat{A}-A} \le \delta_0$ is $O(n^{-2.5})$ if $m>\frac{10A}{c \delta_0^2} n\log n$.
\end{proposition}

\begin{proof}
    By Chernoff bound, for any $0<\epsilon<1$, 
    $$\text{Pr}[|X-\frac{m'cA}{n}|>\frac{\epsilon m'cA}{n}|] < e^{-\frac{\epsilon^2m'cA}{4n}} $$
    Let $\epsilon=\frac{c \delta_0}{A}$, the RHS will be $e^{-\frac{\delta_0^2m}{4cn}} < e^{-2.5 \log n} = O(n^{-2.5})$,
\end{proof}

We now give the algorithm that recovers the order for each of the 2 different choices of functions $f$ provided there are sufficiently many vertices. Specifically, we prove the following two theorems. The probability of success indicated in the theorems is over the randomness of the location of the points as well as the realization of the graph.

\begin{theorem} \label{thm:order-exp}
When $f(x)= e^{-x}$, for any $0<\delta<0.1$ and $m \ge \Theta\left( \frac{n \log n}{c^2 \delta^2} \right)$, there is a poly-time algorithm that recovers the order with probability $1-o(1)$.
\end{theorem}

\begin{theorem} \label{thm:order-lin}
When $f(x) = \frac{1}{x+1}$, for any $0<\delta<0.1$ and $m \ge \Theta\left( \frac{n \log^2 n}{c \delta^2} \right)$, there is a poly-time algorithm that recovers the order with probability $1-o(1)$.
\end{theorem}

The basic idea of both algorithms is that, we first approximate the distance between any pair of vertices. The approximation does not need to be very precise in general -- we only need the precision when the real distance is within a narrow range. When it is outside that range, the approximation only needs to answer that it is out of range. Since we cannot distinguish between a vector of positions and its reflection, we find a vertex that is very close to an endpoint, and assume that that endpoint is 0, the left end of the segment. Then we use the distance approximations to build the relationship between every pair of vertices that are sufficiently far apart. In other words, for each sufficiently distant pair $(u,v)$,  we decide which of $u$ and $v$ is to the left. From these pairwise relationships, we recover the global order.

We  define what we mean by a good approximation of the distance between two vertices.

\begin{definition}
A distance function $d:V \times V \rightarrow \mathbb{R}$ is refered to as a $(L,U,\delta)$-approximation if for any pair of vertices $v_i$ and $v_j$, $d(v_i,v_j)$ satisfies:
\begin{itemize}
\item If $|x_j-x_i|<L$, $d(v_i,v_j)<L+\delta$.
\item If $L \le |x_j-x_i| \le U$, $|x_j-x_i|-\delta < d(v_i,v_j) < |x_j-x_i|+ \delta $
\item If $|x_j-x_i|>U$, $d(v_i,v_j)>U-\delta$.
\end{itemize}
\end{definition}

We say $d$ is a {\em good approximation} if it is an $(L,U,\delta)$-approximation with $3\delta<L<\frac{n}{2}-2\delta$ and $U>2L+8\delta$.  
%\sampath{It seems like the definition of `good' approximation should put an upper bound on $L$... otherwise, we may get an approximation with $L = N/2$ or so, and we won'tt be able to use it to find the order.} 
We present the algorithm that recovers the order given good approximations. We then present algorithms that produce good approximations for each of the probability functions.

\begin{lemma}\label{lem:recalg}
There is an algorithm that recovers the order of the vertices if we are given an $(L,U,\delta)$-approximate distance function with $3\delta<L<\frac{n}{2}-2\delta$ and $U>2L+8\delta$ with probability $1-o(1)$.
\end{lemma}

In Section~\ref{sec:alg}, we describe such an algorithm. We follow this up with good approximation schemes for $f(x)=e^{-x}$ in Section~\ref{sec:exp}, and $f(x) = \frac{1}{x+1}$ in Section~\ref{sec:lin}.

%% file: order_alg.tex
\subsection{Order Recovery from Approximate Distances} \label{sec:alg}

In this section, we give an algorithm (ALGORITHM~\ref{alg:order}) to recover the order of vertices on the segment when we are given 
a $(L,U,\delta)$-approximate distance function $d$ with $3\delta<L<\frac{n}{2}-2\delta$ and $U>2L+8\delta$. The algorithm works as follows: for any triple of vertices $v_i$, $v_j$, and $v_k$, if $v_j$ is in the middle, then the distance between $v_k$ and $v_i$ is larger than $\card{x_i-x_j}$ and $\card{x_j-x_k}$. With a good distance approximation, we can detect which vertex is in the middle, in all triples of vertices that are not too far or too close. We store these ordered triples in a set $S$ (Lemma~\ref{lem:triple}). For any vertex which never occurs in the middle of an ordered triple in $S$, it must be close to one of the endpoints of the segment. Arbitrarily fixing the position of one such vertex as being near the left endpoint, we can `recursively orient' each triple in $S$ (Lemma~\ref{lem:leftright}), which means that we can tell the order of any vertices that are not too close (Lemma~\ref{lem:pair}). Finally, we use this information to give the full order (Lemma~\ref{lem:alg}). Lemma~\ref{lem:recalg} immediately follows from Lemma~\ref{lem:alg}.

\begin{algorithm}[htb]
\caption{Order Recovery}\label{alg:order}
For any pair of points $v_i$ and $v_j$, let $d(v_i,v_j)$ be a $(L,U,\delta)$ approximation of $\card{x_i-x_j}$ with $3\delta<L<\frac{n}{2}-2\delta$ and $U\ge 2L+8\delta$\;
$S \leftarrow \emptyset$ \;
\For {any triple $(v_i,v_j,v_k)$} {
    \If {$d(v_i,v_j) \in [L+\delta,2L+7\delta] \wedge d(v_j,v_k) \in [L+\delta,2L+7\delta] \wedge d(v_i,v_k) > |d(v_i,v_j)-d(v_j,v_k)|+3\delta$}{
        $S \leftarrow S \cup \{(v_i,v_j,v_k)\}$ \;
    }
}
$V' \leftarrow \{v \in V | v \text{ never appears as the middle vertex in any triple in $S$}$\} \;
Pick an arbitrary $v_0 \in V'$\;
$V_0 \leftarrow \{v \in V'| d(v_0,v)>U-\delta\}$\;
$E'=\{(v_i,v_j)|v_i\in V_0 \wedge d(v_i,v_j)\in [L+\delta,2L+7\delta]\}$\;
\While {$S \neq \emptyset$}{
    \For {any triple $(v_i,v_j,v_k)\in S$}{
        \If {$(v_i,v_j)\in E'$}
        {
            $E' \leftarrow E' \cup \{(v_j,v_k)\}$\;
            $S \leftarrow S - \{(v_i,v_j,v_k),(v_k,v_j,v_i)\}$\;
        }
    }
}
Construct a directed graph $G'=(V,E')$ \;
For any vertex $v$, let $R(v)$ be the number of the vertices that can reach $v$ minus the number of vertices reachable from $v$\;
Sort the vertices by $R(v)$ in increasing order and output the order\;
\end{algorithm}

\begin{lemma} \label{lem:triple}
For any triple $(v_i,v_j,v_k)$ in $S$, the location of $v_j$ is in the middle of the location of $v_i$ and $v_k$. On the other hand, for any triple of vertices $(v_i,v_j,v_k)$ such that $v_j$ is in the middle of $v_i$ and $v_k$, $d(v_i,v_j) \in [L+\delta,2L+7\delta]$ and $d(v_j,v_k) \in [L+\delta,2L+7\delta]$, $(v_i,v_j,v_k) \in S$. 
\end{lemma}
\begin{proof}
For any three vertices $v_i$, $v_j$, $v_k$ such that $d(v_i,v_j)$ and $d(v_j,v_k)$ both in $[L+\delta,2L+7\delta]$, we have $\card{x_i-x_j}$ and $\card{x_j-x_k}$ are both between $L$ and $2L+8\delta$ by the definition of $(L,U,\delta)$ approximation. If $v_j$ is in the middle, then $\card{x_i-x_k} \ge d(v_i,v_j)+d(v_j,v_k)-2\delta$, which means $d(v_i,v_k)$ is at least $d(v_i,v_j)+d(v_j,v_k)-3\delta>|d(v_i,v_j)-d(v_j,v_k)|+3\delta$ since both of $d(v_i,v_j)$ and $d(v_j,v_k)$ are at least $L>3\delta$. If $v_j$ is not in the middle, then $\card{x_i-x_k} \le |d(v_i,v_j)-d(v_j,v_k)|+2\delta$, which means $d(v_i,v_k) \le |d(v_i,v_j)-d(v_j,v_k)|+3\delta$. So the triple $(v_i,v_j,v_k)$ is in $S$ if and only if $v_j$ is in the middle.
\end{proof}

By Lemma~\ref{lem:uniform} , for any vertex $v_j$ located between $[L+3\delta,n-L-3\delta]$, there are two vertices $v_i$ and $v_k$ on its left and its right such that $\card{x_i-x_j}$ and $\card{x_j-x_k}$ are both between $L+2\delta,L+3\delta$. This means that $d(v_i,v_j)$ and $d(v_j,v_k)$ are both in $[L+\delta,L+4\delta]$. So $(v_i,v_j,v_k)\in S$ (as $L+4\delta<2L+7\delta$), which implies vertices in $V'$ are located in $[0,L+3\delta]$ or $[n-L-3\delta,n]$. Furthermore, for any vertex pair $(v_i,v_j)$ with $d(v_i,v_j)\in [L+\delta,2L+7\delta]$, there exists a vertex $v_k$ such that $(v_i,v_j,v_k)\in S$ or $(v_k,v_j,v_i)\in S$. Without loss of generality, suppose $v_0 \in [n-L-3\delta,n]$. Then $V_0$ contains all the vertices $v_j$ such that no vertex $v_i$ on its left with $d(v_i,v_j)\in [L+\delta,2L+7\delta]$.

\begin{lemma} \label{lem:leftright}
The while loop of the algorithm always terminates. Moreover, for any pair of vertices $v_i$ and $v_j$, $(v_i,v_j)\in E'$ if and only if $v_i$ is to the left and $d(v_i,v_j)\in [L+\delta,2L+7\delta]$.
\end{lemma}
\begin{proof}
We first prove that for any pair of vertices $(v_i,v_j)$ in $E'$, $v_i$ is to the left of $v_j$, using induction on the order of the pairs added to $E'$. For the base case, $V_0$ only contains vertices with no vertex on their left with approximate distance at least $L+\delta$. So for any pair $(v_i,v_j)$ added into $E'$ before the while loop, $v_i$ is to the left. 
Assume inductively that this is true for all pairs added before the current iteration of the while loop. For any pair $(v_i,v_j)$ added into $E'$ in the current iteration, there is a vertex $v'_i$ such that $(v'_i,v_i,v_j)\in S$ and $(v'_i,v_i)\in E'$. By induction hypothesis, $v'_i$ is on $v_i$'s left. So $v_i$ is between $v'_i$ and $v_j$, so $v_i$ is on $v_j$'s left by Lemma~\ref{lem:triple}. 

We prove that the while loop terminates, i.e., that all triples in $S$ eventually get deleted. Suppose for contradiction that, $v_i$ is the leftmost vertex to appear in any undeleted triple, and there is a triple $(v_i,v_j,v_k)$ that never gets deleted. (Note that whenever $(v_k,v_j,v_i)\in S$, $(v_i,v_j,v_k)\in S$). If there exists a vertex $v'_i$ to the left of $v_i$ with $d(v'_i,v_i)\in [L+\delta,2L+7\delta]$, then $(v'_i,v_i,v_j)$ is in $S$ and will be deleted sometime, then $(v_i,v_j)\in E'$, which means $(v_i,v_j,v_k)$ will be deleted. If there is no such vertex $v'_i$ then $v_i\in V_0$, which also means $(v_i,v_j)\in E'$, $(v_i,v_j,v_k)$ will be deleted in the first iteration. Thus contradicts that $(v_i,v_j,v_k)$ would never gets deleted.

Finally, we prove that any pair of vertices $(v_i,v_j)$ with $d(v_i,v_j)\in [L+\delta,2L+7\delta]$ will be added into $E'$. This is because by Lemma~\ref{lem:uniform} , there exists a vertex $v_k$ such that $(v_i,v_j,v_k) \in S$ or $(v_k,v_j,v_i)\in S$. Since such triple was deleted in the while loop, $(v_i,v_j)$ has been added into $E'$.
\end{proof}

\begin{lemma} \label{lem:pair}
For any pair of vertices $v_i$ and $v_j$, the vertex $v_j$ is reachable from $v_i$ in $G'$ if and only if $d(v_i,v_j)\ge L+\delta$ and $v_i$ is to the left.
\end{lemma}
\begin{proof}
If $v_j$ is reachable from $v_i$, there is a path form $v_i$ to $v_j$, and the location of any vertex on the path is to the left of the next vertex on the path. So $v_i$ is on $v_j$'s left. If $(v_i,v_j)\in E'$, by Lemma~\ref{lem:leftright}, $d(v_i,v_j) \ge L+\delta$, otherwise the path has at least three vertices. By Lemma~\ref{lem:leftright}, any neighbouring vertex has distance at least $L$, which means the distance between $v_i$ and $v_j$ is at least $2L$, so $d(v_i,v_j) \ge 2L-\delta > L+\delta$.

For any pair $v_i$, $v_j$ with $v_i$ to the left and $d(v_i,v_j) \ge L+\delta$, if $d(v_i,v_j) \le 2L+7\delta$, then $(u_i,v_j)\in E'$, which means $v_j$ is reachable from $v_i$ in $G'$. If $d(v_i,v_j)>2L+7\delta$, then the distance between them is at least $2L+6\delta$. by Lemma~\ref{lem:uniform}, there exists a sequece of vertex $v_i=u_1, u_2, \dots, u_k=v_j$ such that for any $1\le \ell \le k-1$, $u_{\ell}$ is to the left of $u_{\ell+1}$, and the distance between them is between $L+2\delta$ and $2L+6\delta$, which means $d(u_{\ell},u_{\ell+1})\in [L+\delta,2L+7\delta]$, in other words, by Lemma~\ref{lem:leftright}, $(u_{\ell},u_{\ell+1})\in E'$, so $v_j$ is reachable form $v_i$ in $G'$. 
\end{proof}

\begin{lemma} \label{lem:alg}
The output order of the algorithm satisfies that for any $v_i$ and $v_j$ that are separated by a distance of at least $3\delta$, $v_i$ appears prior to $v_j$ in the order if and only if $v_i$ is to the left of $v_j$.
\end{lemma}

\begin{proof}
If $v_i$ is to the left and the distance between $v_i$ and $v_j$ is at least $3\delta$, for any vertex $v_k$ on $v_j$'s right with $d(v_j,v_k) \ge L+\delta$, we have $x_k-x_j\ge L$, which means $x_k-x_i \ge L+3\delta$ and $d(v_i,v_k)\ge L+2\delta$. For any vertex $v_k$ on $v_i$'s left with $d(v_i,v_k) \ge L+\delta$, $x_i-x_k \ge L$, which means $x_j-x_k \ge L+3\delta$ and $d(x_k,x_j) \ge L+2\delta$. So $R(x_i)\le R(v_j)$. On the other hand, by Lemma~\ref{lem:uniform} and the fact that $L<\frac{n}{2}-2\delta$, there exists a vertex $v_k$ with one of the following two properties: 
\begin{itemize}
\item $v_k$ is on $v_j$'s right and $x_k-x_j < L$ and $x_k-x_i > L+2\delta$.
\item $v_k$ is on $v_i$'s left and $v_i-v_k < L$ and $v_j-v_k > L+2\delta$.
\end{itemize}
In the first case, $d(v_j,v_k)<L+\delta$ while $d(v_i,v_k)>L+\delta$, which means $v_k$ is reachable from $v_i$ but not $v_j$. In the second case, $d(v_i,v_k)<L+\delta$ while $d(v_j,v_k)>L+\delta$, which means $v_j$ is reachable from $v_k$ but $v_i$ is not reachable from $v_k$.
So $R(v_j)$ is strictly larger than $R(v_i)$.
\end{proof}

%% file: order-exp.tex
\subsection{Distance Approximation for Exponential Decay Function} \label{sec:exp}

%{\bf Sanjeev: We should refer to this as exponential decay function throughout. I fixed the title here but the change needs to be made at other places as well.}

In this section, we consider the case that $f(x) = e^{-x}$. The probability of an edge between two vertices $v_i$ and $v_j$, with locations $x_i$ and $x_j$ respectively, is $c \cdot e^{-|x_i-x_j|}$. We first analyze the degree of each vertex and the number of common neighbors between each pair of vertices.

\begin{lemma} \label{lem:exdeg}
    For any vertex $v_i$  located at position $x_i$ on the segment, if we uniformly sample a vertex $v$ on the segment, then the edge $(v_i,v)$ is present with probability $\frac{c}{n}(2-e^{-x_i}-e^{x_i-n})$. In other words, this is the expected probability of an edge from $v_i$, where the expectation is over the choice of the other endpoint $v$.
\end{lemma}

\begin{proof}
    The probability is the expectation of $e^{-|x_i-x|}$ where $x$ is the location of $v$ which is uniformly sampled on the segment. So the probability is 
    \begin{align*} 
         \int_0^n \frac{c}{n} e^{-|x_i-x|} dx 
        = & \frac{c}{n}\int_0^{x_i} e^{x-x_i} dx + \frac{c}{n}\int_{x_i}^n e^{x_i-x} dx \\
        = & \frac{c(2 - e^{-x_i} - e^{x_i-n})}{n}
    \end{align*}
\end{proof}

\begin{lemma} \label{lem:excom}
    For any two vertices $v_i$ and $v_j$  located at $x_i$ and $x_j$ respectively with $x_i<x_j$, if we uniformly sample a vertex $v$ on the segment, then $v$ is a common neighbor of $v_i$ and $v_j$ with probability $\frac{c^2}{n}((x_j-x_i+1)e^{x_i-x_j} - \frac{1}{2}(e^{x_i+x_j-2n}+e^{-x_i-x_j}))$.
\end{lemma}

\begin{proof}
    Let $p(x)$ be the probability that $v$ is a common neighbor of $v_i$ and $v_j$ where $x$ is the location of $v$, then
    $$
    p(x)=
    \begin{cases}
        c^2 \cdot e^{2x-x_i-x_j}, & \mbox{if } x\le x_i \\
        c^2 \cdot e^{x_i-x_j}, & \mbox{if } x_i<x<x_j \\
        c^2 \cdot e^{x_i+x_j-2x}, & \mbox{if } x\ge x_j
    \end{cases}
    $$
    So the overall probability is
    \begin{align*}
        & \int_0^n \frac{1}{n} p(x) dx \\
        = & \frac{c^2}{n}\int_0^{x_i} e^{2x-x_i-x_j}dx + \frac{c^2(x_j-x_i)}{n} e^{x_i-x_j} + \frac{c^2}{n}\int_{x_j}^n e^{x_i+x_j-2x}dx \\
        = & \frac{c^2(x_j-x_i+1)}{n} e^{x_i-x_j}  - \frac{c^2(e^{-x_i-x_j}+e^{x_i+x_j-2n})}{2n}
    \end{align*}
\end{proof}

By Lemma~\ref{lem:excom}, the number of common neighbors of a pair of vertices ``mostly'' depends on the distance between these two vertices. We use the degree of these two vertices to eliminate the effect of the remaining terms. We first prove that we can check if two vertices are far away.

\begin{lemma} \label{lem:exfar}
    If $m>\frac{2500n \log n}{c^2 \delta^2}$, with probability $1-o(1)$, for any two vertices $v_i$ and $v_j$, (a) if they have no common neighbor, then $\card{x_i-x_j} > 2.5$, and (b) if $\card{x_i-x_j}>n/2$, then they have no common neighbor.
\end{lemma}

\begin{proof}
    If $\card{x_i-x_j} \le 2.5$, then one of $e^{-x_i-x_j}$ and $e^{x_i+x_j-2n}$ is $O(e^{-n})$, without loss of generality, suppose $e^{x_i+x_j-2n}$ is $O(e^{-n})$. Since $-x_i-x_j<-\card{x_i-x_j}$, $e^{-x_i-x_j}<e^{-\card{x_i-x_j}}$. By Lemma~\ref{lem:excom}, the probability that a random sampled vertex be a common neighbor of $v_i$ and $v_j$ is at least $\frac{c^2(\card{x_i-x_j}+0.5)}{n}e^{-\card{x_i-x_j}} > \frac{c^2}{2n}e^{-2.5} > \frac{c^2}{30n}$. Since $m>\frac{2500n \log n}{c^2 \delta^2}$, the probability that $v_i$ and $v_j$ have no common neighbor is $o(n^{-80})$.

    If $\card{x_i-x_j} > n/2$, the probability that a random vertex be a common neighbor of them is at most $e^{-n/2}$. So with probaiblity $1-o(n^{-100})$, they have no common neighbor.
\end{proof}

We now describe how to approximate the distance between two vertices.

%{\bf Sanjeev: We should somewhere (assuming this is true), that whenever we say $1 - o(1)$, we mean $1 - 1/poly(n)$. Also, what is the difference between segment and line? We are using both.}

\begin{lemma} \label{lem:app-exp}
    If $0<\delta<0.1$ and $m>\frac{2500n \log n}{c^2 \delta^2}$, then for any pair of vertices $v_i$ and $v_j$, with probability $1-O(n^{-2.5})$, we can calculate $\hat{d}$, an approximation of $d=\card{x_i-x_j}$ such that:
    \begin{itemize}
        \item If $d<0.3$, $\hat{d}<0.3+\delta$.
        \item If $0.3 \le d \le 2.5$, $d-\delta < \hat{d} < d+ \delta $
        \item If $d>2.5$, $\hat{d}>2.5-\delta$.
    \end{itemize}
\end{lemma}
\begin{proof}
    For any number $x$, let $g(x) = (x+1)e^{-x}$ and $h(x) = e^{-x}+e^{x-n}$. We first prove that we can either approximate $g(d)$ with additive error at most $0.2d$ or directly output a $\hat{d}$ which satisfies the condition. 

    We first check if $v_i$ and $v_j$ have common neighbors. If they have no common neighbor, then by Lemma~\ref{lem:exfar}, $d>2.5$. So we can directly output $\hat{d}=n$. Otherwise we have $d<n/2$.

    By Lemma~\ref{lem:excom} and Proposition~\ref{sam}, we can approximate $g(d) + \frac{1}{2}(e^{x_i+x_j-2n}+e^{-x_i-x_j})$ with additive error $\frac{\delta}{11}$ since $m > \frac{2500n\log n}{c^2\delta^2}$. To eliminate the terms $e^{x_i+x_j-2n}$ and $e^{-x_i-x_j}$, we use the degree of $v_i$ and $v_j$. By Lemma~\ref{lem:exdeg} and Proposition~\ref{sam}, we can approximate $h(x_i)$ and $h(x_j)$ with additive error $\frac{\delta}{11}$. On the other hand, $h(x_i) \cdot h(x_j) = e^{-x_i-x_j} + e^{x_i+x_j-2n} + e^{-n+x_i-x_j} + e^{-n-x_i+x_j}$. The last two terms are $o(1)$ since $\card{x_i-x_j}<n/2$. So we can approximate $e^{-x_i-x_j}+e^{x_i+x_j-2n}$ with additive error $\frac{2\delta}{11} + o(1) < \frac{\delta}{5}$. We can thus approximate $g(d)$ with additive error at most $\frac{\delta}{5}$.

    The proof is completed by the observation that $g(x)$ is monotone decreasing when $x \ge 0$, and the derivative $g'(x)<-0.2$ when $0.3\le x \le 2.5$.
\end{proof}

Note that if $0<\delta<0.1$, $3\delta<0.3<\frac{n}{2}-2\delta$ and $2.5>0.3 \times 2+8\delta$. Theorem~\ref{thm:order-exp} immediately follows from Lemma~\ref{lem:recalg} and Lemma~\ref{lem:app-exp}.

%% file: linear.tex
\subsection{Distance Approximation for Inverse Linear Decaying Function} \label{sec:lin}
In this section, we deal with the case that $f(x) = \frac{c}{x+1}$ and thus the probability of an edge existing between two vertex $v_i$ and $v_j$ with location $x_i$ and $x_j$ on the segment be $\frac{c}{|x_i-x_j|+1}$. We first analyze the degree of each vertex and the number of common neighbors between each two vertices; proofs are deferred to the full version.

\begin{lemma} \label{lem:lindeg}
    Suppose a vertex $v_i$ is located at $x_i$, if we uniformly sample a vertex $v$ on the segment then an edge $(v_i,v)$ will be presented with probability $\frac{c\log(x_i+1)+c\log(n-x_i+1)}{n}$
\end{lemma}

\begin{proof}
    The probability is 
    \begin{align*}
        \frac{c}{n}\int_0^n (|x-x_i|+1)^{-1} dx & = \frac{c}{n} (\int_1^{x_i+1} x^{-1} dx + \int_1^{n-x_i+1} x^{-1} dx) \\
                                                & = \frac{c(\log (x_i+1) + \log(n-x_i+1))}{n}
    \end{align*}
\end{proof}

\begin{lemma} \label{lem:lincom}
    Suppose two vertices $v_i$ and $v_j$ are located at $x_i$ and $x_j$ on the segment with $x_i<x_j$ and $d=x_j-x_i$, if we uniformly sample a vertex $v$ on the segment, then $v$ is a common neighbor of $v_i$ and $v_j$ with probability
    \begin{align*}
    \frac{c^2}{n}\Bigg ( \log (d+1)\left (\frac{2}{d}+\frac{2}{d+2}\right ) +\frac{1}{d}(\log(x_i+1) - \log (x_j+1) 
     + \log (n-x_j+1) - \log (n-x_i+1)) \Bigg)
    \end{align*}
\end{lemma}

\begin{proof}
    The probability is 
    \begin{align*}
        & \frac{c^2}{n} \int_0^n (|x-x_i|+1)^{-1}(|x-x_j|+1)^{-1} dx \\
        = & \frac{c^2}{n} \Bigg ( \int_1^{x_i+1} \frac{1}{x(x+d)} dx + \int_1^{d+1} \frac{1}{x(d+2-x)} dx \\
        &+ \int_1^{n-x_j+1} \frac{1}{x(x+d)} dx \Bigg ) \\
        = & \frac{c^2}{n} \Bigg ( \int_1^{x_i+1} \frac{1}{d}\left (\frac{1}{x} - \frac{1}{(x+d)} \right ) dx + \int_1^{n-x_j+1} \frac{1}{d}\left (\frac{1}{x} - \frac{1}{x+d} \right )dx \\
        &+ \int_1^{d+1} \frac{1}{d+2}\left ( \frac{1}{x}+\frac{1}{(d+2-x)} \right ) dx\Bigg ) \\
        = & \frac{c^2}{n} \Bigg ( \frac{1}{d} (\log (x_i+1) - \log (x_j+1) + \log (n-x_j+1)  \\
        &-\log (n-x_i+1) + 2\log (d+1)) + \frac{1}{d+2} (2 \log (d+1)) \Bigg )
    \end{align*}
\end{proof}

We next show that it can be inferred if a vertex $v_i$ is close to one of the endpoints. If so, we can further approximate its location to within a multiplicative error. In the rest of this section, let $\eps = \frac{\delta}{20}$. 

\begin{lemma} \label{lem:linmult}
    If $m> \frac{40 n \log^2 n} {c \eps^2}$ and $0<\eps<\frac{1}{10}$, then with probability $1-o(1)$, for any vertex $v_i$, we can output a number $\hat{x}_i$ such that:
    \begin{itemize}
        \item if $\bar{x}_i>\frac{9}{\eps}-1$, then $\hat{x}_i>\frac{2}{\eps} + 1$, and
        \item if $\bar{x}_i\le \frac{9}{\eps}-1$, then $\card{\hat{x}_i - \bar{x}_i} < (1+\eps)(\bar{x}_i + 1)$.
    \end{itemize}
    where $\bar{x}_i = \min \{x_i, n-x_i \}$.
\end{lemma}

\begin{proof}
    Since $m>\frac{100n\log^2 n}{c \eps^2}$. By Proposition~\ref{sam} and Lemma~\ref{lem:lindeg}, we can approximate $\log (x_i+1) + \log (n-x_i+1) = \log (\bar{x}_i+1) + \log (n-\bar{x}_i+1)$ within additive error $\frac{\eps}{3}$ with probability $1-o(1)$. Let $a$ be this value, we prove that $\hat{x}_i = e^{a-\log n}-1$ satisfies the requirement.

    $a-\log n = \log (\frac{(\bar{x}_i+1)(n-\bar{x}_i+1)}{n}) \pm \frac{\eps}{3} = \log (\bar{x}_i+1) + \log (1-\frac{\bar{x}_i-1}{n}) \pm \frac{\eps}{3}$. By Proposition~\ref{cla:1-y}, $\log (1-\frac{\bar{x}-1}{n}) = o(1)$ if $\bar{x}_i<\frac{9}{\eps}-1$ and at most $1$ otherwise. 

    If $\bar{x}_i>\frac{9}{\eps}-1$, $a-\log n > \log (\frac{9}{\eps}) - 1 - \frac{\eps}{3} > \log (\frac{3}{\eps}) - \frac{\eps}{3}$. So $\hat{x}_i > (1-\frac{\eps}{2}) \cdot \frac{3}{\eps} - 1 = \frac{3}{\eps} - 2.5 > \frac{2}{\eps} + 1$ since $\eps < \frac{1}{10}$.

    If $\bar{x}_i\le \frac{9}{\eps}-1$, $a-\log n = \log (\bar{x}_i+1) \pm \frac{\eps}{2}$ So $\hat{x}_i + 1 = (1 \pm (e^{\eps/2}))(\bar{x}_i+1) = (1 \pm \eps)(\bar{x}_i+1)$.
\end{proof}

\begin{lemma} \label{lem:app-lin}
    Suppose $0<\delta<0.1$ and $m>\frac{16000 n \log^2 n}{c \delta^2}$, with probability $1-o(1)$, for any two vertex $v_i$ and $v_j$ with distance $d$, we can approximate $d$ by $\hat{d}$ which satisfies:
    \begin{itemize}
        \item $\hat{d}<d+\delta$ if $d<0.3$.
        \item $d-\delta<\hat{d}<d+\delta$ if $0.3\le d \le 2$.
        \item $\hat{d}>d-\delta$ if $d>2$.
    \end{itemize}
\end{lemma}

\begin{proof}
    For any number $a$, $b$, denote $g(a,b) = \frac{\log a - \log b}{a-b}$ and $h(a) = \log (a+1) ( \frac{2}{a}+\frac{2}{a+2} )$. We first prove that we can either approximate $h(d)$ with additive error at most $2\eps$ or directly output a $\hat{d}$ which satisifies the condition.
    By Lemma~\ref{lem:lincom} and Proposition~\ref{sam}, we can approximate $h(d) - g(x_i+1,x_j+1) - g(n-x_i+1,n-x_j+1)$ with additive error $\frac{\eps}{c \sqrt{\log n}} = o(1)$, Denote $a$ as this value.
     
    Let $\hat{x}_i$ and $\hat{x}_j$ be the value given by Lemma~\ref{lem:linmult}. If $\hat{x}_i$ and $\hat{x}_j$ are both at least $\frac{1}{\eps}$, then $v_i$ and $v_j$ are both at least $\frac{1}{\eps}-1$ far away from both endpoints. By the argument in the proof of Proposition~\ref{prop:linear}, $g(x_i+1,x_j+1)$ and $g(n-x_i+1,n-x_j+1)$ are both at most $\eps$. So $\card{a-h(d)} < 2\eps$. If one of $\hat{x}_i$ and $\hat{x}_j$ larger than $\frac{2}{\eps}+1$ and the other less than $\frac{1}{\eps}$, then $\card{x_j-x_i} > \frac{2}{\eps} - (1+\eps)\frac{1+\eps}{\eps} > 2$. So we can directly output $\hat{d}=n$. The only case remaining is when both of $\hat{x}_i$ and $\hat{x}_j$ at most $\frac{2}{\eps}+1$. 

    In this case, $x_i$ and $x_j$ are both at most $\frac{3}{\eps}$ far away from one of the endpoint. If they are close to different endpoint, then $d > n/2$, which menas $\expect{a} = O(\frac{1}{n})$ and $a = o(1)$. Otherwise $\expect{a} = \Omega(1) - o(1)$ and thus $a = \Omega(1)$. So we can check if $v_i$ and $v_j$ are close to the same endpoint. If not, $x_j-x_i>n/2$ and so we can directly output $\hat{d}=n$. Then we focus on the case that they are close to the same endpoint. Without loss of generality, suppose both of $x_i$ and $x_j$ are at most $\frac{3}{\eps}$.

    If $\hat{x}_i$ and $\hat{x}_j$ are both at most $8$, then both of $x_i$ and $x_j$ are at most $9(1+\eps)-1 < 9$, which means $\card{\hat{x}_i-x_i}$ and $\card{\hat{x}_j-x_j}$ are both at most $10\eps = \frac{\delta}{2}$. Then we can output $\hat{d}=\card{\hat{x}_i-\hat{x}_j}$. If one of $\hat{x}_i$ and $\hat{x}_j$ is at least $8$ and the other is at most $5$, then $\card{x_i-x_j} > 3 (1-2\eps) >2$. So we can output $\hat{d}=n$. The only case remaining is when both of $\hat{x}_i$ and $\hat{x}_j$ are at least $5$. In this case, $x_i$ and $x_j$ are both larger than $4$. By Proposition~\ref{prop:linear}, $\card{g(x_i,x_j)-g(\hat{x}_i,\hat{x}_j)} < \eps$. So $a-g(\hat{x}_i,\hat{x}_j)$ is an approximation of $h(d)$ with additive error at most $\eps + o(1) < 2\eps$.

    By this point, we either already output a $\hat{d}$ which satisfies the condition or have an approximation of $h(d)$ with additive error $2\eps$. To complete the proof we observe that the function $h(d)$ is monotone decreasing when $d > 0$ and that the derivative of $h(d)$ is strictly less than $-0.1$ when $0.5 \le d \le 2$.
\end{proof}

Note that if $0<\delta<0.1$, $3\delta<0.5<\frac{n}{2}$ and $2>0.5+8\delta$. Theorem~\ref{thm:order-lin} immediately follows from Lemma~\ref{lem:recalg} and Lemma~\ref{lem:app-lin}.

%% file: position.tex
\section{Recovering the Position} \label{sec:position}

In this section, we consider the problem of recovering the positions of the vertices on the segment. First, we prove the following simple result, which extends the results for recovering the order.

\begin{theorem} \label{lem:ordpos}
Suppose $m>\frac{10n^2\log n}{\delta^2}$. For any function $f$, if we can recover the order of the vertices, then we can also recover a position vector $\hat{X}$ such that for any $i$, $\card{x_i-\hat{x}_i} < 2 \delta$ with probability $1-o(1)$.
\end{theorem}

\begin{proof}
    Suppose the order output by the order recovery algorithm is $(v_1,v_2,\dots,v_m)$, and their true positions are $(x_1,x_2,\dots,x_m)$. We will prove that $\card{x_i-\frac{in}{m}}<2\delta$ (i.e. we can just output the position as uniformly dispersed along the segment according to the order). 

    Suppose the real order is $(u_1,u_2,\dots,u_m)$, and the real positions are $(y_1 < y_2 < \cdots < y_m)$. We first prove $\card{x_i-y_i}<\delta$, and then prove that $\card{y_i-\frac{in}{m}} <\delta$. The following arguments are based on the event that the run of the order recovery algorithm is successful.

    For any $i$, if $x_i-y_i\ge \delta$, then for any $j \le i$, $x_i-y_j \ge \delta$. By the definition of recovering the order, for any $j\le i$, $u_j$ occurs before $v_i$ in the order output by the algorithm, which contradicts the fact that $v_i$ appears at the $i^{th}$ position of the order output by the algorithm. So $x_i-y_i<\delta$. For the same reason, we also have $y_i-x_i<\delta$.

    On the other hand, for any $1 \le k \le \frac{2n}{\delta}$, let $Z_k$ be the number of vertices sampled in segment $[0,k\delta/2]$. By the Chernoff bound, with probability $1-o(\frac{1}{n})$, $\card{Z_k-\frac{km\delta}{2n}} < \frac{m}{2\delta n}$. By taking the union bound over the complementary events, all $Z_k$'s are close to their expectation with probability $1-o(1)$. For any $i$, suppose $\frac{(k-1)m\delta}{2n} < i \le \frac{km\delta}{2n}$, then there are at most $i$ vertices sampled in the segment $[0,(k-2)\delta/2]$ and at least $i$ vertices sampled in the segment $[0,(k+1)\delta/2]$, which implies $(k-2)\delta/2 < y_i < (k+1) \delta/2$. On the other hand, $(k-1)\delta/2 < i \le k \delta/2$, so $\card{y_i-\frac{in}{m}} < \delta$.
\end{proof}

By Theorem~\ref{lem:ordpos} and the results in Section~\ref{sec:order}, we can recover the position with $\tilde{\Omega}(n^2)$ vertices for both choices of $f$. However, there is a huge gap compared to the number of samples necessary for recovering the order. 
\\[1em]
\noindent
{\bf \large Sample complexity of identifying best position vector}
In the remainder of this section, we consider the following ``weaker'' problem: the task is distinguishing two position vectors $X$ and $Y$ where $X=(x_1,x_2,\dots,x_m)$ and $Y=(y_1,y_2,\dots,y_m)$ with the guarantee that vertices in $X$ and $Y$ have the same order. We focus on the exponential decay function $f(x)=e^{-x}$ and the case when the number of samples is between the gap of Theorem~\ref{thm:order-exp} and Theorem~\ref{lem:ordpos}. We say that two position vectors $X$ and $Y$ are {\em $\delta$-far} if there exists a vertex $v_i$ such that $\card{x_i-y_i}>\delta$. We prove that we cannot distinguish two positions which are $\delta$ far away when there are $o(n^{1.5})$ samples. This shows that we cannot recover the position of vertices with only $o(n^{1.5})$ samples even if the algorithm is given the order.

\begin{theorem} \label{thm:lb}
    For any $m <\frac{0.05n^{1.5}}{\delta}$, if $X$ is sampled uniformly at random, then with probability $1-o(1)$, we can construct a position vector $Y$ which has the same order as $X$ and is $\delta$-far from $X$ such that, for any tester $\Psi$ that determines whether a graph is generated from $X$ or $Y$, if we randomly select a postion vector $Z$ from $\{X,Y\}$, and sample a graph $G$ according to $Z$, there is a constant probability that $\Psi(G)\neq Z$. 
\end{theorem}

On the other hand, we prove that if $m=\Omega(n^{1.5} \log n)$, then we can distinguish any two position vectors which are far from each other when one vector is sampled uniformly, which means Theorem~\ref{thm:lb} is tight up to a $O(\log n)$ factor. 

\begin{theorem} \label{thm:ub}
    For any $\frac{n^{1.5} \log n}{\delta}<m < n^2$, if $X$ is sampled uniformly at random, then with probability $1-o(1)$, for any position vector $Y$ with the same vertex order as $X$ and $\delta$-far from $X$, suppose we randomly sample a graph $G$ according to $X$, then with probability $1-o(1)$, $P_X(G)>P_Y(G)$.
\end{theorem}

We prove Theorem~\ref{thm:lb} in Section~\ref{sec:lb}, and prove Theorem~\ref{thm:ub} in Section~\ref{sec:ub}.

%% file: lb.tex
\subsection{Proof of Theorem~\ref{thm:lb}} \label{sec:lb}
For any tester $\Psi$ which decides whether a graph $G$ is generated from $X$ or $Y$, let $P_X(\Psi(G) \neq X)$ (resp. $P_Y(\Psi(G) \neq Y)$) be the probability that $X$ (resp. $Y$) generates a graph $G$ such that $\Psi(G)=Y$ (resp. $\Psi(G)=X$). By Le Cam's method \cite{le2012asymptotic,yu1997assouad}, we have
\begin{align*}
    P_X(\Psi(G) \neq X) + P_Y(\Psi(G) \neq Y) \ge 1 - \tvd{P_X}{P_Y}
\end{align*}
where $\tvd{P_X}{P_Y}$ is the total variation distance between $P_X$ and $P_Y$. On the other hand, by Pinsker’s inequality,
\begin{align*}
    \tvd{P_X}{P_Y} \le \sqrt{\frac{1}{2}\kl{P_X}{P_Y}}
\end{align*}
where $\kl{P_X}{P_Y}$ is the Kullback–Leibler divergence between $P_X$ and $P_Y$. Therefore, to prove that any tester $\Psi$ cannot distinguish $X$ and $Y$, we only need to prove $\kl{P_X}{P_Y}$ is bounded away from $2$. By definition,
\begin{align*}
    \kl{P_X}{P_Y} &= \sum_G P_X(G) (\log P_X(G) - \log P_Y(G)) \\
                  &= \Exp_{G \sim X} \bracket{\log P_X(G) - \log P_Y(G)}
\end{align*}

From this point, we use $\Exp$ to simplify $\Exp_{G \sim X}$. Denote $L = \log P_X(G) - \log P_Y(G)$, and $L_{i,j} = \log (e^{-\card{x_i-x_j}}) - \log (e^{-\card{y_i-y_j}})$ if $(v_i,v_j) \in G$ and $L_{i,j} = \log (1-e^{-\card{x_i-x_j}}) - \log (1-e^{-\card{y_i-y_j}})$ if $(v_i,v_j) \notin G$. Again by definition, 

$$  \kl{P_X}{P_Y} = \expect{L} = \sum_{i,j} \expect{L_{i,j}}.$$

\iffalse
\begin{align*}
    \kl{P_X}{P_Y} = \expect{L} = \sum_{i,j} \expect{L_{i,j}} \\
\end{align*}
\fi

Now we define the location vector $Y$ that confuses the tester. Without loss of generality, suppose $x_1 < x_2 < \dots < x_m$. Let $Y$ be the position vector $(y_1,y_2,\dots,y_m)$ such that $y_i = (1-\frac{2\delta}{n}) x_i$. It is easy to see that as long as $m$ is super constant, $\card{x_m-y_m}>\delta$ with probability $1-o(1)$, which means $X$ and $Y$ are $\delta$-far. To proof Theorem~\ref{thm:lb}, we only need to prove $\expect{L} = 2-\Omega(1)$.

Throughout this section, we let $d_{i,j} = \card{x_i-x_j}$ and $d'_{i,j}$ as $\card{x_i-x_j}-\card{y_i-y_j}$.
The following lemma gives the upper bound on $\expect{L_{i,j}}$.

\begin{lemma} \label{lem:lbXij}
    For any pair of vertices $v_i,v_j$, 
%    $$e^{-d_{i,j}}(d'^2_{i,j}/2+(1-e^{-d_{i,j}})a/2) < \expect{L_{i,j}} < e^{-d_{i,j}}(d'^2_{i,j}+\frac{2d'^2_{i,j}}{d_{i,j}})$$
    $$\expect{L_{i,j}} < e^{-d_{i,j}}(d'^2_{i,j}+\frac{2d'^2_{i,j}}{d_{i,j}})$$
\end{lemma}

\begin{proof}
    By definition of $L_{i,j}$, with probability $e^{-d_{i,j}}$, $L_{i,j}=-d'_{i,j}$ and with probability $1-e^{-d_{i,j}}$, $L_{i,j}=\log (1-e^{-d_{i,j}}) - \log (1-e^{-d_{i,j}+d'_{i,j}}) = - \log (\frac{1-e^{-d_{i,j}+d'_{i,j}}}{1-e^{-d_{i,j}}})$. So 
    \begin{align*}
    \expect{L_{i,j}} &=-d'_{i,j}e^{-d_{i,j}}-(1-e^{-d_{i,j}})\log (\frac{1-e^{-d_{i,j}+d'_{i,j}}}{1-e^{-d_{i,j}}}) \\
                     &= -d'_{i,j}e^{-d_{i,j}}-(1-e^{-d_{i,j}})\log (1- \frac{e^{-d_{i,j}}(e^{d'_{i,j}}-1)}{1-e^{-d_{i,j}}})
    \end{align*}
    by Proposition~\ref{cla:log}, $a=\frac{e^{-d_{i,j}}(e^{d'_{i,j}}-1)}{1-e^{-d_{i,j}}} < \frac{d'_{i,j}}{d_{i,j}} < 0.5$. 
    Together with Proposition~\ref{cla:1-y},
    \begin{align*}
        \expect{L_{i,j}} &<-d'_{i,j}e^{-d_{i,j}} + e^{-d_{i,j}}(e^{d'_{i,j}}-1)(1+a) \\
                         &< e^{-d_{i,j}}(e^{d'_{i,j}}-d'_{i,j}-1+\frac{d'_{i,j}(e^{d'_{i,j}}-1)}{d_{i,j}}) 
    \end{align*}
    Since $d'_{i,j}<1/2$, $e^{d'_{i,j}}<1+d'_{i,j}+d'^2_{i,j}$ and $e^{d'_{i,j}}<1+2d'_{i,j}$, which means $\expect{L_{i,j}} < e^{-d_{i,j}}(d'^2_{i,j}+2d'^2_{i,j}/d_{i,j})$.

\iffalse
    Again by Proposition~\ref{cla:1-y},
    \begin{align*}
    \expect{L_{i,j}} &> -d'_{i,j}e^{-d_{i,j}} + e^{-d_{i,j}}(e^{d'_{i,j}}-1)(1+a/2) \\ 
                     &> e^{-d_{i,j}}(e^{d'_{i,j}}-d'_{i,j}-1+(e^{d'_{i,j}}-1)a/2) \\ 
                     &> e^{-d_{i,j}}(d'^2_{i,j}/2 + (e^{d'_{i,j}}-1)a/2)
    \end{align*}
\fi
\end{proof}

Now we give an upper bound on $\expect{L}$.

\begin{lemma} \label{lem:ubex}
    If $m<\frac{0.05n^{3/2}}{\delta}$ and $X$ is obtained by sampling each point uniformly, then $\expect{L} = \expect{\sum_{i,j} L_{i,j}} < 1$ with probability $1-o(1)$.
\end{lemma}

\begin{proof}
    Let the $S_1,S_2,\dots,S_n$ be the set of vertices where $S_k$ contains all the vertices inside the interval $[i,i+1]$ in $X$. Let $i$, $j$ be two vertices inside $S_k$ and $S_{\ell}$ where $k\le \ell$, then $\expect{L_{i,j}} \le 6(\ell-k+1)^2e^{-(\ell-k-1)}\cdot \frac{\delta^2}{n^2}$ by Lemma~\ref{lem:lbXij} and the fact that the distance between $i$ and $j$ is at least $\ell-k-1$ and at most $\ell-k+1$, $\card{y_i-y_j} = (1-\frac{2\delta}{n}) \card{x_i-x_j}$. So
    \begin{align*}
        \expect{\sum_{i,j} L_{i,j}} & = \sum_{k,\ell} \sum_{i \in S_k, j\in S_{\ell}} \expect{L_{i,j}} \\
                                    & \le \frac{\delta^2}{n^2}\sum_{k,\ell} |S_k|\cdot |S_{\ell}| 6(\ell-k+1)^2e^{-(\ell-k-1)} \\
                                    & = \frac{\delta^2}{n^2}\sum_{k=0}^{n-1} \sum_{\ell=1}^{n-k} |S_{\ell}| \cdot |S_{\ell+k}| 6(k+1)^2 e^{-(k-1)}
    \end{align*}
    By Rearrangement inequality \cite{zygmund1953hg}, for any $k$, $\sum_{\ell=1}^{n-k} |S_{\ell}| \cdot |S_{\ell+k}| \le \sum_{\ell=1}^n |S_{\ell}|^2$. So
    \begin{align*}
        \expect{\sum_{i,j} L_{i,j}} & \le \frac{\delta^2}{n^2}(\sum_{k=1}^n |S_k|^2) \cdot (\sum_{k=0}^{n-1} 6(k+1)^2 e^{-(k-1)}) \\
                                    & \le \frac{\delta^2}{n^2}(6e + \sum_{k=0}^\infty (6k^2+24k+24) e^{-k)})(\sum_{k=1}^n |S_k|^2) \\
                                    & \le \frac{\delta^2}{n^2}(6e + \frac{6e(1+e)}{(e-1)^3} + \frac{24e}{(e-1)^2} + \frac{24e}{e-1}) \cdot (\sum_{k=1}^n |s_k|^2) \\                   
                                    & \le \frac{100\delta^2}{n^2}\sum_{k=1}^n |S_k|^2
    \end{align*}
    By the choice of $m$, each $|S_k|<2m/n < \frac{0.1n^{1/2}}{\delta} $ with probability $1-o(1)$ by Chernoff bound, so $\sum_{k=1}^n |S_k|^2 \le \frac{10 ^ {-2} n^2}{\delta^2}$, which means $\expect{\sum_{i,j} L_{i,j}} < 1$.
\end{proof}

%% file: ub.tex
\subsection{Proof of Theorem~\ref{thm:ub}} \label{sec:ub}

We define $L_{i,j}$ and $L$ the as in Section~\ref{sec:lb}. To prove Theorem~\ref{thm:ub}, we need to prove $\prob{L>0}=1-o(1)$. The basic idea is to prove $\expect{L}$ is large and use the concentration bound (Propostion~\ref{prop:Hof}) to prove $\expect{L}$ is larger than the ``concentration range''.

The main difficulty is that since the location vector $Y$ is chosen adversarily, some $L_{i,j}$'s might be ``ill-behaved'' and thus their deviation is hard to control due to the choice of $Y$. To solve this problem, we construct $\bar{L}_{i,j}$ as follows: If $\card{y_i-y_j}>\card{x_i-x_j}$, then let $\bar{L}_{i,j}=\min \{2,L_{i,j}\}$ if $(v_i,v_j)\in G$; if $\card{y_i-y_j}<\card{x_i-x_j}$, then let $\bar{L}_{i,j} = (1 - e^{-L_{i,j}})+\frac{1}{2}(1 - e^{-L_{i,j}})^2$; if $(v_i,v_j) \notin G$. In any scenerio, $\bar{L}_{i,j}$ is always smaller than $L_{i,j}$. (This is due to Proposition~\ref{cla:1-y}.) So $\prob{\sum_{i,j} \bar{L}_{i,j} > 0} \le \prob{L>0}$. Moreover, let $\bar{L}$ be the sum of $\bar{L}_{i,j}$ excluding those pairs $i,j$ where $\card{x_i-x_j} > 5 \log n$ and $\card{x_i-x_j} > \card{y_i-y_j}$. For such pairs, the probability that $(v_i,v_j) \notin G$ is $1-O(n^{-5})$ and in that event, $\bar{L}_{i,j}>0$. Since there are at most $m^2 = o(n^5)$ pairs of such $i,j$, with probability $1-o(1)$, all of these $\bar{L}_{i,j}$'s are greater than $0$. So with probability $1-o(1)$, $\bar{L} \le \sum_{i,j} \bar{L}_{i,j} \le L$. So it is sufficient to prove $\prob{\bar{L} > 0} = 1 - o(1)$. We call the unexcluded pairs as the pair contributing to $\bar{L}$.
Throughout this section, let $d_{i,j} = \card{x_i-x_j}$ and $d'_{i,j} = \card{\card{x_i-x_j}-\card{y_i-y_j}}$.

We first prove a simple lemma about the distance between each pair of vertices in $X$.

\begin{lemma} \label{lem:ubfar}
If $m = \tilde{O}(n^2)$, with probability $1-o(1)$, for any pair $i,j$, $\card{x_i-x_j}>\frac{1}{n^4}$.
\end{lemma}

\begin{proof}
For any pair $i,j$, the probability that $\card{x_i-x_j} \le \frac{1}{n^4}$ is at most $\frac{(2/n^4)}{n} = O(\frac{1}{n^5})$. Since there are at most $m^2 = o(n^5)$ pairs, so with probability $1-o(1)$ there is no such pair.
\end{proof}

Hereafter, we assume $d_{i,j} > \frac{1}{n^4}$ for all pair of $i,j$. We establish the following property of $\bar{L}_{i,j}$.

\begin{lemma} \label{lem:ubsubexp}
For any pair $i,j$ that contributes to $\bar{L}$, $\bar{L}_{i,j}$ is a sub-exponential random variable with parameter $(\sigma_{i,j},b)$ where $\sigma_{i,j}^2 = 10 \log n \cdot \expect{\bar{L}_{i,j}}$ and $b = 10 \log n$.
\end{lemma}

\begin{proof}
By Proposition~\ref{cla:subexp}, it is sufficient to prove that for any $\lambda<\frac{1}{b}$, 
$$
\expp{\bar{L}_{i,j},\bar{L}'_{i,j}}{e^{\frac{\lambda^2(\bar{L}_{i,j}-\bar{L}'_{i,j})^2}{2}}} < e^{\frac{\lambda^2\sigma_{i,j}^2}{2}}
$$
where $\bar{L}'_{i,j}$ is a random varibale independent and identical to $\bar{L}_{i,j}$.
We prove the lemma respectively in the case of $\card{y_i-y_j} \le \card{x_i-x_j}$ and $\card{y-i-y_j} < \card{x_i-x_j}$.

\textbf{Case 1:} $\card{y_i-y_j} \le \card{x_i-x_j}$. Denote $a=\frac{e^{-d_{i,j}}(e^{d'_{i,j}}-1)}{1-e^{-d_{i,j}}}$. $\bar{L}_{i,j} = -d'_{i,j}$ with probability $e^{-d_{i,j}}$ and $a + \frac{1}{2} a^2$ with probability $(1-e^{-d_{i,j}})$. So 
\begin{align*}
\expect{\bar{L}_{i,j}} &= -d'_{i,j}e^{-d_{i,j}} + (1-e^{-d_{i,j}})(a+\frac{1}{2} a^2) \\
                       &= e^{-d_{i,j}}(e^{d'_{i,j}}-d_{i,j}-1) + \frac{1}{2} (1-e^{-d_{i,j}})a^2 \\
                       &\ge \frac{1}{2}(e^{-d_{i,j}} d'^2_{i,j} + (1-e^{-d_{i,j}}) a^2)
\end{align*}
So $e^{\frac{\lambda^2\sigma^2}{2}} > 1 + 5\lambda^2 e^{-d_{i,j}} (d'^2_{i,j} +a^2) \log n$.

On the other hand, $(\bar{L}_{i,j}-\bar{L}'_{i,j})^2 = (d'_{i,j}+a)^2 \le 2d'^2_{i,j}+2a^2$ with probability $2e^{-d_{i,j}}(1-e^{-d_{i,j}}$ and $0$ otherwise. By the condition that $\bar{L}_{i,j}$ contributes to $\bar{L}$, $d'_{i,j} \le d_{i,j} \le 5 \log n$; by Proposition~\ref{cla:log}, $a \le 1$. So $\lambda^2(2d'^2_{i,j}+2a^2) < \frac{50 \log^2 n + 2}{100\log ^2 n} < 1$ for any $\lambda<\frac{1}{b}$. Which means
\begin{align*}
\expp{\bar{L}_{i,j},\bar{L}'_{i,j}}{e^{\frac{\lambda^2(\bar{L}_{i,j}-\bar{L}'_{i,j})^2}{2}}} 
    &\le 1 + \expp{\bar{L}_{i,j},\bar{L}'_{i,j}}{\lambda^2(\bar{L}_{i,j}-\bar{L}'_{i,j})^2} \\
    &\le 1 + 2e^{-d_{i,j}}(1-e^{-d_{i,j}})(2d'^2_{i,j}+2a^2)\lambda^2
\end{align*}
which means
$$
\expp{\bar{L}_{i,j},\bar{L}'_{i,j}}{e^{\frac{\lambda^2(\bar{L}_{i,j}-\bar{L}'_{i,j})^2}{2}}} < e^{\frac{\lambda^2\sigma_{i,j}^2}{2}}
$$

\textbf{Case 2:} $\card{y_i-y_j} > \card{x_i-x_j}$. Denote $a = \log (\frac{1-e^{-(d_{i,j}+d'_{i,j})}}{1-e^{-d_{i,j}}})$. $\bar{L}_{i,j} = \min \{d'_{i,j},2\}$ with probability $e^{-d_{i,j}}$ and $-a$ with probability $(1-e^{-d_{i,j}})$. Since $d_{i,j} \ge \frac{1}{n^4}$, $\log (1-e^{-d_{i,j}}) \ge \log (1-e^{-n^{-4}}) \ge \log (\frac{1}{2n^4}) \ge -5\log n$, which means $a<5\log n$. So $\lambda^2(\bar{L}_{i,j}-\bar{L}'_{i,j})^2 \le (2a^2+2)\lambda^2 < 1$ for any $\lambda<\frac{1}{10\log n} = \frac{1}{b}$. So
$$
\expp{\bar{L}_{i,j},\bar{L}'_{i,j}}{e^{\frac{\lambda^2(\bar{L}_{i,j}-\bar{L}'_{i,j})^2}{2}}} \le 1 + \expp{\bar{L}_{i,j},\bar{L}'_{i,j}}{\lambda^2(\bar{L}_{i,j}-\bar{L}'_{i,j})^2} 
$$
On the other hand, $e^{\frac{\lambda^2\sigma^2}{2}} > 1 + 5\lambda^2\expect{\bar{L}_{i,j}} \log n$, so we just need to prove 
$$\expp{\bar{L}_{i,j},\bar{L}'_{i,j}}{(\bar{L}_{i,j}-\bar{L}'_{i,j})^2} \le 5\expect{\bar{L}_{i,j}} \log n$$

\textbf{Case 2.1:} If $d'_{i,j} \ge 2$, 
\begin{align*}
\expp{\bar{L}_{i,j},\bar{L}'_{i,j}}{(\bar{L}_{i,j}-\bar{L}'_{i,j})^2} 
    &= 2e^{-d_{i,j}}(1-e^{-d_{i,j}})(8+2a^2) \\
    &< 16e^{-d_{i,j}} + 4(1-e^{-d_{i,j}})a \log n \\
    &< 16e^{-d_{i,j}} + 4e^{-d_{i,j}}(1-e^{-d'_{i,j}}) \log n \\
    &< 5e^{-d_{i,j}} \log n
\end{align*}
On the other hand, $\expect{\bar{L}_{i,j}} = 2 e^{-d_{i,j}} - (1-e^{-d_{i,j}})a > e^{-d_{i,j}} (2 - (1-e^{-d'_{i,j}})) > e^{-d_{i,j}}$, so $5\expect{\bar{L}_{i,j}} \log n > 5e^{-d_{i,j}} \log n$. 

\textbf{Case 2.2:} If $d'_{i,j} \le d_{i,j}$ and $d'_{i,j} < 2$,
$\expp{\bar{L}_{i,j},\bar{L}'_{i,j}}{(\bar{L}_{i,j}-\bar{L}'_{i,j})^2} < 2e^{-d_{i,j}}(1-e^{-d_{i,j}})(2d'^2_{i,j}+2a^2) $. Let $z = \frac{e^{-d_{i,j}}(1-e^{-d'_{i,j}})}{1-e^{-d_{i,j}}}$, by Proposition~\ref{cla:log2}, $z<\frac{d'_{i,j}}{d_{i,j}} \le 1$, so $a = \log (1+z) < z - \frac{z^2}{2} + \frac{z^3}{3} < z-\frac{z^2}{6}$, which means $\expect{\bar{L}_{i,j}} > e^{-d_{i,j}}d'_{i,j} - (1-e^{-d_{i,j}})(z-\frac{z^2}{6}) = e^{-d_{i,j}}(d'_{i,j}+e^{-d'_{i,j}}-1)+\frac{1}{6}z^2(1-e^{-d_{i,j}})$ where $e^{-d'_{i,j}}+d'_{i,j}-1 > \frac{d'^2_{i,j}}{2} - \frac{d'^3_{i,j}}{6} > \frac{d'^2_{i,j}}{6}$ since $d'_{i,j} < 2$. So $\expect{\bar{L}_{i,j}} > \frac{1}{6}e^{-d_{i,j}}d'^2_{i,j} + \frac{1}{6}z^2(1-e^{-d_{i,j}})$. On the other hand, $2e^{-d_{i,j}}(1-e^{-d_{i,j}})(2d'^2_{i,j}+2a^2)<4e^{-d_{i,j}}d'^2_{i,j}+4(1-e^{-d_{i,j}})z^2$. So $\expp{\bar{L}_{i,j},\bar{L}'_{i,j}}{(\bar{L}_{i,j}-\bar{L}'_{i,j})^2} < \expect{\bar{L}_{i,j}} \cdot \log n$.

\textbf{Case 2.3:} If $d_{i,j} < d'_{i,j} < 2$, let $\eps=e^{-d_{i,j}} \frac{d'_{i,j}}{d_{i,j}}$ and $z = \frac{e^{-d_{i,j}}(1-e^{-d'_{i,j}})}{1-e^{-d_{i,j}}}$. Since $a=\log (1+z) < z$, $\expect{\bar{L}_{i,j}} > e^{-d_{i,j}}(d'_{i,j}- 1 + e^{-d'_{i,j}}) > \frac{1}{6} e^{-d_{i,j}} d'^2_{i,j}$ since $d'_{i,j}<2$. On the other hand, since $d'_{i,j}>d_{i,j}$, $\frac{1-e^{-d'_{i,j}}}{1-e^{-d_{i,j}}} < \frac{d'_{i,j}}{d_{i,j}}$ by Proposition~\ref{cla:log3}, so $a< \log (1+e^{-d_{i,j}}\frac{d'_{i,j}}{d_{i,j}}) = \log (1+\eps)$, and $\expect{\bar{L}_{i,j}} > d_{i,j} \eps - (1-e^{-d_{i,j}}) \log (1+\eps) > d_{i,j} (\eps - \log (1+\eps)) > \frac{d_{i,j}}{2}\log (1+\eps)^2$ (the last inequality is due to $e^a-a-1>\frac{a^2}{2}$ for any $a>0$). So
\begin{align*}
\expect{\bar{L}_{i,j}} > \frac{1}{2}(\frac{1}{6}e^{-d_{i,j}} d'^2_{i,j} + \frac{d_{i,j}}{2} \log(1+\eps)^2) 
                       > \frac{1}{24}(e^{-d_{i,j}}(1-e^{-d_{i,j}}) ( d'^2_{i,j} + 3 a^2) )
\end{align*}
which means
\begin{align*}
\expp{\bar{L}_{i,j},\bar{L}'_{i,j}}{(\bar{L}_{i,j}-\bar{L}'_{i,j})^2} \le 4e^{-d_{i,j}}(1-e^{-d_{i,j}})(d'^2_{i,j} + a^2) 
                                                                      \le 5\expect{\bar{L}_{i,j}} \log n
\end{align*}

\end{proof}
Next, we analyze the expectation of $\bar{L}$.
The following lemma is a byproduct  of the proof of Lemma~\ref{lem:ubsubexp}.

\begin{lemma} \label{lem:lbLij}
For any $i$, $j$, $\expect{\bar{L}_{i,j}} > \frac{1}{6} e^{-d_{i,j}} d'^2_{i,j}$ if $d'_{i,j} \le 2$. Otherwise $\expect{\bar{L}_{i,j}} > e^{-d_{i,j}}$.
\end{lemma}

We next show that $d'_{i,j}$ satisfies the triangle inequality.

\begin{lemma} \label{lem:ubtri}
For any $i$, $j$ and $k$, $d'_{i,j} \le d'_{i,k}+d'_{k,j}$.
\end{lemma}

\begin{proof}
Since $X$ and $Y$ has the same vertex order, $d'_{i,k}+d'_{k,j} = \card{x_i-x_k-y_i+x_k} + \card{x_j-x_k-y_j+y_k} \ge \card{x_i-x_j-y_i+y_j} = d'_{i,j}$.
\end{proof}

We prove a lower bound on the expectation of $\bar{L}$.

\begin{lemma} \label{lem:ubexp}
For any $\frac{100n^{1.5} \log n}{\delta} < m < n^2$, if $X$ is sampled uniformly, then with probability $1-o(1)$, for any $Y$ such that there is a pair $i,j$ with $d'_{i,j}> \frac{\delta}{2}$,  $\expect{\bar{L}}>5 \log^2 n$.
\end{lemma}

\begin{proof}
By Lemma~\ref{lem:lbLij}, $\expect{\bar{L}_{i,j}} \ge 0$. It is sufficient to prove that sum of some $\expect{\bar{L}_{i,j}}$ contributed to $\bar{L}$ is larger than $5 \log n$. We first prove that if there is a pair $i'$ and $j'$ satisfies $d_{i',j'} \le 1$ and $d'_{i',j'}>\frac{\delta}{8}$, then $\expect{\bar{L}}>5\log^2 n$. By Chernoff bound, with probability $1-o(1)$ there are at least $\frac{90\sqrt{n} \log n}{\delta}$ vertices in each segment of length $1$. So there are at least $\frac{90\sqrt{n}\log n}{\delta}$ vertices which is at most $1$ away from both $v_{i'}$ and $v_{j'}$. Suppose $v_k$ is such a vertex, then either $d'_{i',k}$ or $d'_{k,j'}$ is at least $\frac{\delta}{16}$ by Lemma~\ref{lem:ubtri}, which means either $\expect{\bar{L}_{i',k}}$ or $\expect{\bar{L}_{k,j'}}$ is at least $\frac{\delta^2}{256e}$ by lemma~\ref{lem:lbLij}. So $\bar{L}>\frac{90\sqrt{n}\log n}{\delta} \cdot \frac{\delta^2}{256e} > 5 \log^2 n$. 

For any integer $K$, let $S_{K}$ be the set of vertex in segment $[K-1,K]$. Let $v_i \in S_I$ and $v_j \in S_J$. Without loss of generality, suppose $I \le J$. Then for any vertex $v_k$ in $S_I$ (resp. $S_j$), if $d'_{i,k}$ (resp. $d'_{k,j}$) is at least $\frac{\delta}{8}$, which means $\expect{L}>5 \log^2 n$. Otherwise, we have $I<J$ and for any $v_k \in S_I$ and $v_{\ell} \in S_J$, $d'_{k,\ell}> \frac{\delta}{4}$. 

For any $I \le K \le J$, let $v_{k_{K}}$ be an arbitrary vertex in $S_K$. We prove that $\sum_{I\le K < J} \expect{\bar{L}_{k_K,k_{K+1}}} \ge \frac{\delta^2}{1000n}$. For any $K$, since $v_{k_K}$ and $v_{k_{K+1}}$ are in $S_K$ and $S_{K+1}$ respectively, $d_{k_K,k_{K+1}} \le 2$, which means $e^{-d_{k_K,k_{K+1}}} > e^{-2} > \frac{1}{10}$. 

If there exists a $K$ such that $d'_{k_K,k_{K+1}} > 2$, then $\bar{L}_{k_K,k_{K+1}} > \frac{1}{10} > \frac{\delta^2}{1000n}$ by Lemma~\ref{lem:ubexp}. Otherwise $\sum_{I\le K < J} \expect{\bar{L}_{k_K,k_{K+1}}} \ge \frac{1}{60} \sum_{I \le K < J} d'^2_{k_K,k_{k+1}}$ by Lemma~\ref{lem:ubexp}. 

Since $d'_{k_I,k_J}>\frac{\delta}{4}$, $\sum_{I \le K <j} d'_{k_K,k_{k+1}} \ge \frac{\delta}{4}$ by Lemma~\ref{lem:ubtri}. By Cauchy-Schwarz inequality, 
$$
\sum_{I \le K <j} d'^2_{k_K,k_{k+1}} \ge \frac{1}{J-I} (\sum_{I \le K <j} d'_{k_K,k_{k+1}})^2 \ge \frac{\delta^2}{16(J-I)} \ge \frac{\delta^2}{16n}
$$
which means $\sum_{I\le K < J} \expect{\bar{L}_{k_K,k_{K+1}}} \ge \frac{\delta^2}{1000n}$.

Let $N=\frac{90\sqrt{n}}{\delta}$ and for any $I \le K \le J$, let $v_{\ell^K_1},v_{\ell^K_2},\dots,v_{\ell^K_N}$ be arbitrary $N$ vertices in $S_k$. Then
\begin{align*}
\expect{\bar{L}} &\ge \sum_{I \le K <J} \sum_{i'=1}^N \sum_{j'=1}^N \expect{\bar{L}_{\ell^K_{i'}, \ell^{K+1}_{j'}}}  \\
                 &= \sum_{I \le K <J} \sum_{i'=1}^N \sum_{j'=0}^{N-1} \expect{\bar{L}_{\ell^K_{i'}, \ell^{K+1}_{(i'+j') \mod N + 1 }}}\\
                 &= \sum_{i'=1}^N \sum_{j'=0}^{N-1} \sum_{I \le K < J} \expect{\bar{L}_{\ell^K_{(i' + Kj') \mod N +1}, \ell^{K+1}_{(i'+(K+1)j') \mod N + 1 }}}\\
                 &\ge \sum_{i'=1}^N \sum_{j'=0}^{N-1} \frac{\delta^2}{1000n} = \frac{N^2 \delta^2}{1000n} > 5 \log^2 n
\end{align*}
\end{proof}

Now we are ready to use the concentration bound (Proposition~\ref{prop:Hof}) to prove Theorem~\ref{thm:ub}.

\begin{proof} [Proof of Theorem~\ref{thm:ub}]
Let $v_j$ (resp. $v_k$) be the left (resp. right) most vertex in $X$, then with probability $1-o(1)$ $x_j = o(1)$ and $x_k = n-o(1)$. Let $v_i$ be the vertex such that $\card{x_i-y_i} > \delta$, then either $d'_{i,j}>\delta-o(1)$ or $d'_{i,k}>\delta-o(1)$. Suppose $d'_{i,j}>\delta-o(1)>\frac{\delta}{2}$. By Lemma~\ref{lem:ubexp}, $\expect{\bar{L}}>5 \log^2 n$. By Lemma~\ref{lem:ubsubexp} and Proposition~\ref{prop:Hof},
\begin{align*}
\prob{\bar{L}<0} &\le \prob{\card{\bar{L} - \expect{\bar{L}}} > \expect{\bar{L}}} \\
                 &< 2 e^{-\frac{\expect{\bar{L}}^2}{20\expect{\bar{L}} \log n}} = 2 e^{-\frac{\expect{\bar{L}}}{20 \log n}} < 2e^{-1.25\log n} \\
                 &= o(1)
\end{align*}
\end{proof}

\vspace*{-0.3in}

%% file: exper.tex
\section{Empirical Results}
\label{empirical}

In this section, we present results on simulating our algorithms on synthetically generated test sets. We created 5 test sets where each set contains $30$ independently generated graphs. In each test set, the points are generated uniformly at random on the line with length $n=25$. The number of points $m$ in the test sets range from $10,000$ to $20,000$. We focus on the probability function $f(x) = e^{-x}$ for the probability of generating edges in each test set.

We run our algorithm in Section~\ref{sec:order} to recover the order of points as well as output the position by the algorithm given in Theorem~\ref{lem:ordpos}. We then analyze the observed error in using our algorithms for recovering the order and recovering the position.

\begin{figure}[htbp]
    \centering
    \centering
    \subcaptionbox{Sample size vs. distance between inverted pairs
    }[0.48 \textwidth]{
        \centering
        \includegraphics[width=3.5in]{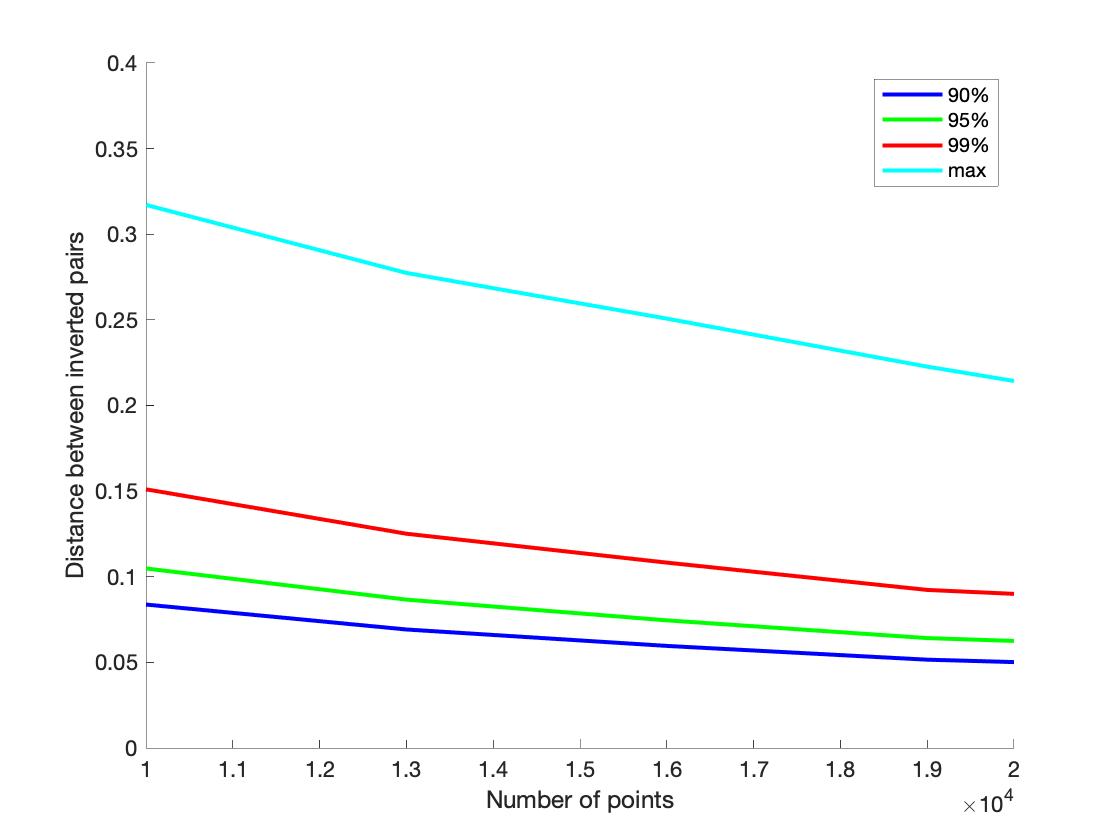}
        %\caption{fig1}
    }
    \hspace{2mm}
    \subcaptionbox{Sample size vs. error in the recovered position
    }[0.48 \textwidth]{ 
    \centering
    \includegraphics[width=3.5in]{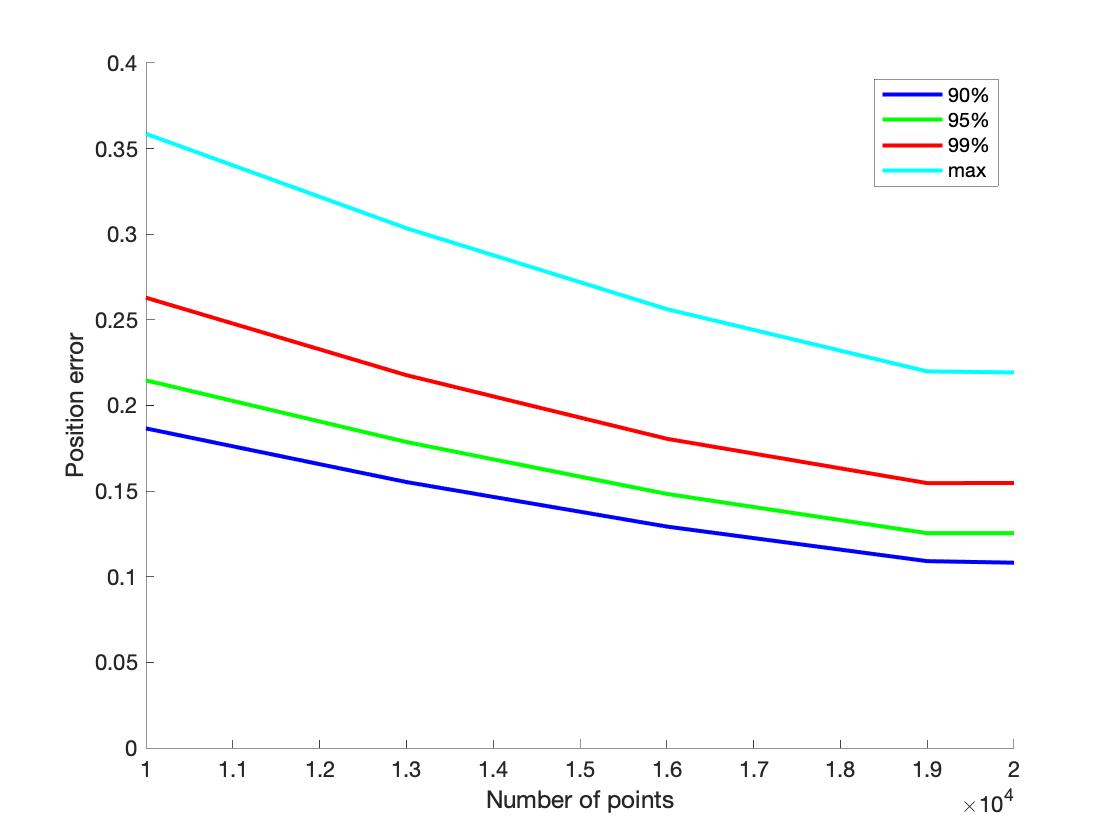}
    }
    \caption{Results}\label{fig:result}
\end{figure}

For the task of recovering the order, we collect all pairs of vertices $(v_i,v_j)$ such that our algorithm outputs them in inverted order. We calculate the distance between each inverted pair, and consider the 90th percentile, 95th percentile, 99th percentile and the maximum distance values among the inverted pairs. For each of these values, we use the average among the 30 tests in each test set. The results are illustrated in Figure~\ref{fig:result}(a). As indicated by our theoretical analysis, as the sample size increases, distance between inverted pairs decreases. For instance, the green line shows that once the sample size exceeds 10000, more than 95\% of inversions occur among pairs that are less than 0.1 distance apart (i.e. very close). 

\iffalse
\begin{figure}[htbp] 
    \centering
    \includegraphics[width=6in]{order2.jpg} 
    \caption{Sample size vs. distance between inverted pairs}
    \label{fig:order} 
\end{figure} 
\fi

For the task of recovering the positions, we calculate the distance between the position output by our algorithm and the actual position for each point. Again, we look the 90th percentile, 95th percentile, 99th percentile and the maximum, and use the average among the 30 tests in each test set. The results are illustrated in Figure~\ref{fig:result}(b). 

\iffalse
\begin{figure}[htbp] 
    \centering
    \includegraphics[width=6in]{position2.jpg} 
    \caption{Sample size vs. error in the recovered position}
    \label{fig:position} 
\end{figure} 
\fi

%% file: conclusions.tex
\section{Conclusions}
\label{sec:conclusions}

We developed a framework for recovery that uses the following high-level approach: 1) use the graph to reconstruct approximate degrees and common neighborhood sizes for pairs of vertices;  2) use this information to approximately identify the neighborhoods of each vertex, and spatial relationships between vertices in each neighborhood; and finally, 3) use the local knowledge to establish global structure - order relations or positions. Using this framework, we obtained essentially tight bounds on the number of samples required for recovering the (approximate) order of points on a line segment under both exponential decay and linear decay models. It would be interesting to close the  gap that remains between the upper and lower bounds for recovering the location of the points. We also empirically analyzed recovery accuracy of our algorithms on synthetic data sets. 

This paper can be seen as taking the first step in what should be a promising line of research, that will include generalizing our results to other metric spaces as well as to other edge probability functions. As we move from one-dimensional space to higher dimensional spaces, recovery becomes distinctly harder (as one might expect) but our preliminary investigation suggests that the framework described in this work continues to be of value in understanding recovery in $\mathbb{R}^{k}$ for $k \ge 2$.  Beyond this, a particularly intriguing  problem is to recover missing attributes. If we are given a graph as well as some partial information about the attributes of vertices, can we learn both the edge probability function and values of the missing attributes? Such problems are likely to be of interest in social science research, as well as in understanding diverse networks such as biological and economic networks.